\documentclass[11pt,dvips]{article}
\usepackage{amsfonts}
\usepackage{graphicx}
\usepackage{amsmath}
\usepackage{amssymb}
\usepackage{latexsym}
\usepackage{color}

\input{tcilatex}
\begin{document}

\thispagestyle{empty}

\begin{center}
\vspace{1.8cm}

%%%%%%%%%%%%%%%%%%%%%%%%%%%%%%%%%%%%%%%%%%%%%%%%%%%%%%%%%%%%%%%%%%%%%%%%%%%%%%%%%%%%%%%%%%%%%%%%%%%%%%%%%%%%%%%%%%%%%%%
{\Large \textbf{\ Non-classical correlations in a two-mode optomechanical
system}}

%%%%%%%%%%%%%%%%%%%%%%%%%%%%%%%%%%%%%%%%%%%%%%%%%%%%%%%%%%%%%%%%%%%%%%%%%%%%%%%%%%%%%%%%%%%%%%%%%%%%%%%%%%%%%%%%%%%%%

\vspace{1.5cm}

\textbf{J. El Qars}$^{a}${\footnote{%
email: \textsf{j.elqars@gmail.com}}}, \textbf{M. Daoud}$^{b,c,d}${\footnote{%
email: \textsf{m$_-$daoud@hotmail.com}}} and \textbf{R. Ahl Laamara}$^{a,e}$ 
{\footnote{%
email: \textsf{ahllaamara@gmail.com}}}

\vspace{0.5cm}

$^{a}$\textit{LPHE-MS, Faculty of Sciences, University Mohammed V, Rabat,
Morocco}\\[0.5em]
$^{b}$\textit{Max Planck Institute for the Physics of Complex Systems,
Dresden, Germany}\\[0.5em]
$^{c}$\textit{Abdus Salam International Centre for Theoretical Physics,
Miramare, Trieste, Italy}\\[0.5em]
$^{d}$\textit{Department of Physics, Faculty of Sciences, University Ibnou
Zohr, Agadir, Morocco}\\[0.5em]
$^{e}$\textit{Centre of Physics and Mathematics (CPM), University Mohammed
V, Rabat, Morocco}\\[0.5em]

\vspace{3cm}\textbf{Abstract}
\end{center}

\baselineskip=18pt \medskip

\noindent The pairwise quantum correlations in a tripartite optomechanical
system comprising a mechanical mode and two optical modes are analyzed. The
Simon criterion is used as a witness of the separability. Whereas, the
Gaussian discord is employed to capture the quantumness of correlations.
Both entanglement and Gaussian discord are evaluated as functions of the
parameters characterizing the environment and the system (temperature,
squeezing and optomechanical coupling). We work in the resolved-sideband
regime. We show that it is possible to reach simultaneous three bipartite
entanglements via the quantum correlations transfer from the squeezed light
to the system. While, even without squeezed light, the quantumness of
correlations can be captured simultaneously between the three modes for a
very wide range of parameters. Specifically, we find that the two optical
modes exhibit more quantum correlations in comparison with the entangled
mechanical-optical modes. Finally, unlike the two hybrid subsystems, the
purely optical one seems more resilient against the environmental
destructive effects.

%%%%%%%%%%%%%%%%%%%%%%%%%%%%%%%%%%%%%%%%%%%%%%%%
\newpage

\section{Introduction}

In quantum information science, the study of quantum correlations is a key
issue. In fact, the entanglement property in multipartite quantum systems is
a fundamental resource for various quantum tasks \cite{Nielsen}. In this
context, quantifying quantum correlations has been the subject of extensive
investigation during the last two decades. A particular attention was
dedicated to entangled states of continuous variables systems, especially
Gaussian states. Indeed, motivated by the experimental implementation and
control of this kind of quantum states, a complete qualitative and
quantitative characterization of the non-classical properties was obtained
in the literature \cite%
{Simon,Duan,Vidal,MGVT,Giedke(1),Adesso(1),M.G.Paris(1)}. Moreover, the
separability of two modes continuous variables (CVs) can be completely
characterized using the Simon criterion \cite{Simon}. We note that now it is
clearly understood that separable states, especially mixed ones, might also
contain quantum correlations and the separability is not an indicator of
classicality. Thus, quantum discord has drawn much attention in recent years
as the most used quantifier to capture the \textit{quantumness} of
correlations in discrete variables systems \cite{Zurek-Vedral} as well as
continuous variables (CVs) \cite{Adesso(2),Giorda}. In either Markovian or
non-Markovian regimes, a considerable efforts have been devoted to
investigate both entanglement and the Gaussian quantum discord in many
different models \cite{Tesfa-Doukas-Izar-Mazzola(1),Mauro(1),IP,Farace}.
Essentially, it has been shown that unlike entanglement, quantum discord is
more robust against the effects of the environment and interestingly, it is
immune to sudden death \cite%
{Mauro(1),IP,Mauro(2)-M.G.Paris(2)-Davidovich,Olivares(1)}.

Encoding quantum information in quantum states of any realistic system
encounters both the quantum decoherence and dissipation induced by the
unavoidable coupling with its environment. In addition, a fairly good
understanding of how to control the coupling between the quantum systems and
their own environment, will make the exploitation of quantum properties for
quantum information processing more effective. In this sense, there has been
considerable interest in studying both decoherence and dissipation process,
which are fundamental issues in quantum physics to understand the transition
between classical and quantum worlds \cite{Weis-Eisert(1),Cirac-Zoller}.

Recently, significant efforts has been deployed towards the macroscopic
quantum state by developing various schemes for their experimental
production. In this context, quantum optomechanical systems constitute a
promising candidate to investigate quantum mechanical effects \cite%
{Marquardt}. Proposals include the ground state optical cooling of the
fundamental mechanical mode \cite{Mari(1),Clerk(1),Meystre}, the creation of
macroscopic quantum superpositions or so-called Schr\"{o}dinger's cat states 
\cite{Bose-Marshall}, quantum state transfer \cite{Braunstein}, the
detection of the gravitational waves \cite{Braginski-Mertz}, entangling
states of mechanical modes to each other \cite%
{Mauro(1),Eisert(2)-Tombisi(1),Pinard,G.Agarwal} or optical modes \cite%
{Vitali(1),Grebogi-Eisert(3)}, the entanglement generation between two
optical modes \cite{Daoud,Giovanitti(1)}. Now, it is more or less accepted
that the encoding information in optomechanical systems can constitute a
promising candidate in the field of quantum information science.

We consider an optomechanical setup where a movable mirror is placed inside
a Fabry-Perot cavity. The analysis of the entanglement in such tripartite
optomechanical system, where a single mechanical mode is coupled to two
optical cavities modes via radiation pressure, was studied in Ref \cite%
{Mauro(3)}, knowing that no further approximation has been done apart from
the linearization around the classical steady state. The aim of the present
work is to go one step further. Indeed, feeding the same optomechanical
system by two-mode squeezed light and using the rotating wave approximation
(RWA), we shall study the quantum correlations behavior between the
different modes of the system. We use the Simon criterion as a witness of
the separability \cite{Simon} and we will extend our analyses far beyond
entanglement trying to detect \textit{the quantumness} of pairwise
correlations in three different subsystems using the Gaussian quantum
discord \cite{Adesso(2),Giorda}.

The organization of this paper is as follows. In section \ref{sec2}, we
introduce the basic model, give the quantum Langevin equations describing
the dynamics of the single mechanical mode and the two optical modes. The
needed approximations to get the explicit form of the covariance matrix are
also discussed. In section \ref{sec3}, using the Simon criterion, the
bi-separability between any pair of modes is studied in terms of the
temperature to understand the thermal effects on the entanglement properties
of the system. We also investigate the behavior of the entanglements under
both the squeezing and the optomechanical coupling effects. In section \ref%
{sec4}, we investigate the quantum correlations in the system far beyond
entanglement. For this, the pairwise Gaussian quantum discord among the
three bipartite subsystems are computed and analyzed. Concluding remarks
close this paper.

\section{Formulation and theoretical description of the system \label{sec2}}

\subsection{Model}

\begin{figure}[tbh]
\centerline{\includegraphics[width=8.5cm]{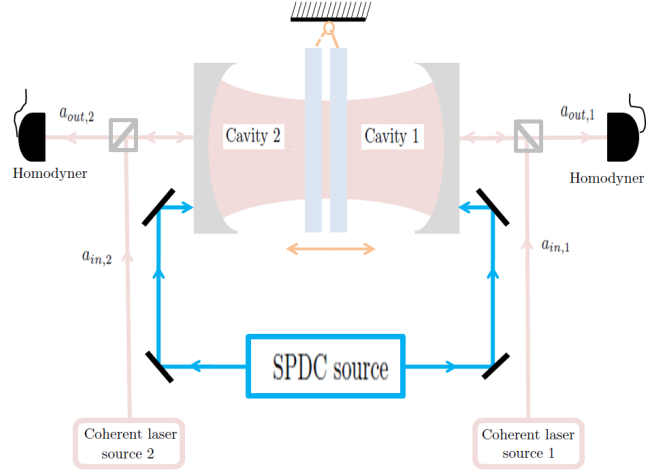}}
\caption{Schematic description of a double-cavity optomechanical system. The
movable mirror which represented as a mass on a pendulum with the mechanical
frequency $\protect\omega _{\protect\mu }$, the damping rate $\protect\gamma %
_{\protect\mu }$ and an effective mass $m_{\protect\mu }$ is modeled as a
quantum mechanical harmonic oscillator. The system is injected
simultaneously by two-mode squeezed light and coherent laser fields trough
the two partially transmitting mirrors. In addition, $a_{in,j}$
(respectively. $a_{out,j}$) represents the $j^{\mathrm{th}}$ input
(respectively. the $j^{\mathrm{th}}$ output) laser field with $j=1,2$. By
means of the homodyner systems, it is possible to evaluate numerically $%
a_{out,1}$ and $a_{out,2}$ which allows us to obtain the correlation matrix
of the global system. Subsequently, one can compute both the entanglement
and the Gaussian quantum discord in different bipartite subsystems.}
\label{double-cavity}
\end{figure}

The system under study, illustrated in Fig. \ref{double-cavity}, is a
Fabry-Perot double-cavity system comprising one movable perfectly reflecting
mirror, which is inserted between two fixed partially transmitting mirrors.
The movable mirror is coupled simultaneously by means of radiation pressure
to the right (respectively. left) optical cavity mode of frequency $\omega
_{c_{1}}$ (respectively. $\omega _{c_{2}}$). Each optical cavity mode
(labled as $o_{j}$ for $j=1,2$) is driven by an external coherent laser
source with the input power $P_{j}$, phase $\varphi _{j}$ and frequency $%
\omega _{L_{j}}$. In addition, we assume that the system is pumped by
two-mode squeezed light produced for example by spontaneous parametric
down-conversion source (SPDC) \cite{Burnham-Shih}. The first (respectively,
second) squeezed mode is sent towards the right (respectively. left) side
cavity. Finally, the movable mirror will be modeled as a quantum mechanical
harmonic oscillator having an effective mass $m_{\mu }$, a mechanical
frequency denoted by $\omega _{\mu }$ and a mechanical damping rate $\gamma
_{\mu }$.

In a frame rotating at the frequency $\omega _{L_{j}}$ ($j=1,2$) of the
lasers, the Hamiltonian describing the optomechanical system under
consideration can be written as \cite{Law} 
\begin{equation}
H=\omega _{\mu }b^{\dag }b+\sum_{j=1}^{2}\left( (\omega _{c_{j}}-\omega
_{L_{j}})a_{j}^{\dag }a_{j}+(-1)^{j}g_{j}a_{j}^{\dag }a_{j}(b^{\dag
}+b)+\varepsilon _{j}(a_{j}^{\dag }e^{i\varphi _{j}}+a_{j}e^{-i\varphi
_{j}})\right) .  \label{Hamil}
\end{equation}
\noindent As mentioned above, the movable mirror will be treated as a single
mechanical mode (labeled as $m$) defined by the annihilation and creation
operators $b,b^{\dag }$ with $\left[ b,b^{\dag }\right] =1$. We denote by $%
a_{j}$ and $a_{j}^{\dag }$ the annihilation and creation operators of the $%
j^{th}$ optical cavity mode with $\left[ a_{j},a_{k}^{\dag }\right] =\delta
_{jk}$ (for $j,k=1,2$). The optomechanical single-photon coupling rate $%
g_{j} $ between the mechanical mode and the $j^{th}$ optical cavity mode is
given by $g_{j}=\left( \omega _{c_{j}}/l_{j}\right) \left( \hbar /m_{\mu
}\omega _{\mu }\right) ^{1/2}$. The quantities $l_{j}$ stands for the $%
j^{th} $ cavity length (with $j=1,2$). While, the coupling strength between
the $j^{th}$ external laser and its corresponding cavity field is defined by 
$\varepsilon _{j}=\left( 2\kappa _{j}P_{j}/\hbar \omega _{L_{j}}\right)
^{1/2} $ $($for $j=1,2)$, where $\kappa _{j}$ referring to the energy decay
rate of the $j^{th}$ cavity.

\subsection{Quantum Langevin equations}

In the Heisenberg picture, the dynamics of the mechanical mode and the $%
j^{th}$ optical cavity mode is completely described by the following set of
nonlinear quantum Langevin equations 
\begin{eqnarray}
\partial _{t}b &=&-\bigg(\frac{\gamma _{\mu }}{2}+i\omega _{\mu }\bigg)%
b+\sum_{j=1}^{2}(-1)^{j+1}ig_{j}a_{j}^{\dag }a_{j}+\sqrt{\gamma _{\mu }}%
b^{in},  \label{b_dot} \\
\partial _{t}a_{j} &=&-\bigg(\frac{\kappa _{j}}{2}-i\Delta _{j}\bigg)%
a_{j}+(-1)^{j+1}ig_{j}a_{j}\left( b^{\dag }+b\right) -i\varepsilon
_{j}e^{i\varphi _{j}}+\sqrt{\kappa _{j}}a_{j}^{in}\text{ \ for \ \ }j=1,2
\label{cj_dot}
\end{eqnarray}%
where $\Delta _{j}=\omega _{L_{j}}-\omega _{c_{j}}$ (for $\ j=1,2$) is the $%
j^{th}$ laser detuning \cite{Marquardt}. Moreover, $b^{in}$ is the random
Brownian operator with zero-mean value ($\langle b^{in}\rangle =0$)
describing the coupling of the movable mirror with its own environment. In
general, $b^{in}$ is not $\delta $-correlated \cite{Giovanitti(2)}. However,
quantum effects are reached only using oscillators with a large mechanical
quality factor $\mathcal{Q}=\omega _{\mu }/\gamma _{\mu }\gg 1$, which
allowing us to recover the Markovian process. In this limit, we have the
following nonzero time-domain correlation functions \cite{P.Zoller} 
\begin{eqnarray}
\langle b^{in}(t)b^{in\dag }(t^{\prime })\rangle &=&(n_{\mathrm{th}%
}+1)\delta (t-t^{\prime }),  \label{ther_noise_1} \\
\langle b^{in\dag }(t)b^{in}(t^{\prime })\rangle &=&n_{\mathrm{th}}\delta
(t-t^{\prime }),\text{\ \ }  \label{ther_noise_2}
\end{eqnarray}%
where $n_{\mathrm{th}}=\bigg(\exp (\hbar \omega _{\mu }/k_{B}T)-1\bigg)^{-1}$
is the mean thermal photons number, $T$ is the temperature of the mirror
environment and $k_{B}$ is the Boltzmann constant. Another kind of noise
affecting the system is the $j^{th}$ input squeezed vacuum noise operator $%
a_{j}^{in}$ ($j=1,2$) with zero mean value. They satisfy the following non
zero time-domain correlation properties given by \cite%
{Mauro(1),G.Agarwal,Gardiner} 
\begin{eqnarray}
\langle a_{j}^{in}(t)a_{j}^{in\dag }(t^{\prime })\rangle &=&(N+1)\delta
(t-t^{\prime })\text{ \ }j=1,2\text{ ,\ }  \label{S_1} \\
\langle a_{j}^{in\dag }(t)a_{j}^{in}(t^{\prime })\rangle &=&N\delta
(t-t^{\prime })\text{ \ \ \ \ \ \ \ }j=1,2\text{ ,}  \label{S_2} \\
\langle a_{j}^{in\dag }(t)a_{k}^{in\dag }(t^{\prime }) &=&Me^{i\omega _{\mu
}(t+t^{\prime })}\delta (t-t^{\prime })\text{\ \ }k\neq j=1,2\text{ ,}
\label{S_3} \\
\langle a_{j}^{in}(t)a_{k}^{in}(t^{\prime }) &=&Me^{-i\omega _{\mu
}(t+t^{\prime })}\delta (t-t^{\prime })\text{ \ \ \ }k\neq j=1,2\text{ , }
\label{S_4}
\end{eqnarray}%
with $N$ $=$ $\mathrm{sinh}^{\mathrm{2}}r$, $M$ $=$ $\mathrm{sinh}r\mathrm{%
cosh}r$ where $r$ being the squeezing parameter.

\subsection{ Linearization of quantum Langevin equations}

Due to the nonlinear nature of the radiation pressure, the investigation of
the exact quantum dynamics of the whole system is non trivial, which
subsequently makes impossible to get a rigorous analytical solutions of Eqs.
(\ref{b_dot}) and (\ref{cj_dot}). To overcome this difficulty, we adopt the
linearization approach discussed in \cite{G.J.Milburn-Fabre}. Indeed, in
order to attain satisfactory levels of optomechanical interaction which
leads to a stationary and robust entanglement, the two cavities should be
intensely driven by strong power lasers so that the intra-cavities fields
are strong. Assuming this, the steady-state mean value of each bosonic
operator is larger in comparison with the corresponding fluctuation, i.e., $%
\left\vert \langle a_{j}\rangle \right\vert =$ $\left\vert a_{js}\right\vert
\gg \left\vert \delta a_{j}\right\vert $ for $j=1,2$ and $\left\vert \langle
b \rangle \right\vert =\left\vert b_{s}\right\vert \gg \left\vert \delta
b\right\vert $. In this sense, we consider the dynamics of small
fluctuations around the steady state of the system by decomposing each
operator ($b$, $a_{1}$ and $a_{2}$) into two parts, i.e., sum of its mean
value and a small fluctuation with zero mean value ($\langle \delta
a_{j}\rangle =0$ for $j=1,2$ and $\langle \delta b\rangle =0$), so 
\begin{equation}
b=b_{s}+\delta b\ \ ,\ \ a_{j}=a_{js}+\delta a_{j}\ \ \text{for \ }j=1,2%
\text{ ,}  \label{Linearisation}
\end{equation}%
\noindent where the mean values $b_{s}$ and $a_{js}$ (for $j=1,2$) are
complex-numbers and can be evaluated by setting the time derivatives to zero
and factorizing the averages in Eqs. (\ref{b_dot}) and (\ref{cj_dot}). Thus,
the steady-state values $b_{s}$ and $a_{js}$ write as 
\begin{eqnarray}
b_{s} &=&\frac{2i}{\gamma _{\mu }+2i\omega _{\mu }}\sum%
\limits_{j=1}^{2}(-1)^{j+1}g_{j}\left\vert a_{js}\right\vert ^{2}\text{ , }
\label{aver_b} \\
a_{js} &=&-2ie^{i\varphi _{j}}\frac{\varepsilon _{j}}{\kappa _{j}-2i\Delta
_{j}^{\prime }}\text{\ for\ \ }j=1,2\text{ ,}  \label{aver_cj}
\end{eqnarray}%
where $\Delta _{j}^{\prime }$ $=$ $\Delta _{j}$ $+(-1)^{j+1}g_{j}(b_{s}^{%
\ast }+b_{s})$ is the $j^{th}$ effective cavity detuning including the
radiation pressure effects \cite{Marquardt,Tombisi(2)}. When the two
cavities are intensely driven so that the intracavities fields are strong,
i.e., $\left\vert a_{js}\right\vert \gg 1$ for $j=1,2$, the nonlinear terms $%
\delta a_{j}^{\dag }\delta a_{j}$, $\delta a_{j}\delta b$ and $\delta
a_{j}\delta b^{\dag }$ (for $j=1,2$), can be safely neglected. Hence, one
gets the following linearized Langevin equations 
\begin{eqnarray}
\ \text{\ \ }\ \text{ \ }\partial _{t}\delta b &=&-\left( i\omega _{\mu }+%
\frac{\gamma _{\mu }}{2}\right) \delta b+\sum_{j=1}^{2}(-1)^{j}\mathcal{G}%
_{j}\left( \delta a_{j}-\delta a_{j}^{\dag }\right) +\sqrt{\gamma _{\mu }}%
b^{in},  \label{delta_b_dot} \\
\partial _{t}\delta a_{j} &=&-\left( \frac{\kappa _{j}}{2}-i\Delta
_{j}^{\prime }\right) \delta a_{j}+(-1)^{j+1}\mathcal{G}_{j}\left( \delta
b^{\dag }+\delta b\right) \ +\sqrt{\kappa _{j}}a_{j}^{in}\text{ \ \ for \ \ }%
j=1,2\text{ }.\   \label{delta_cj_dot}
\end{eqnarray}

\noindent In the two last equations, the parameter $\mathcal{G}_{j}$ (for $%
j=1,2$) defined by $\mathcal{G}_{j}$ $=g_{j}\left\vert a_{js}\right\vert
=g_{j}\sqrt{\bar{n}_{\mathrm{cav}}^{j}}$ is the $j^{th}$ light-enhanced
optomechanical coupling for the linearized regime \cite{Marquardt}. While,
the quantity $\bar{n}_{\mathrm{cav}}^{j}$ represents the number of photons
circulating inside the $j^{th}$ cavity \cite{Marquardt}. Using the bosonic
linearization (\ref{Linearisation}), the Hamiltonian (\ref{Hamil}) can be
rewritten as 
\begin{equation}
H_{eff}=\omega _{\mu }\delta b^{\dag }\delta b+\sum_{j=1}^{2}\left[
(-1)^{j+1}\Delta _{j}^{\prime }\delta a_{j}^{\dag }\delta a_{j}+i(-1)^{j}%
\mathcal{G}_{j}(\delta a_{j}-\delta a_{j}^{\dag })(\delta b+\delta b^{\dag })%
\right],  \label{H-eff-1}
\end{equation}
where $\mathcal{G}_{j}$ can be interpreted as a new $j^{th}$ effective
optomechanical coupling. In fact, it is simple to see that the linearized
quantum Langevin equations (\ref{delta_b_dot}) and (\ref{delta_cj_dot}) can
be directly obtained from the effective Hamiltonian (\ref{H-eff-1}). One can
verify that the operators $\delta a_{j}$ and $\delta b$ satisfy the usual
bosonic commutations rules (i.e. [$\delta a_{j}$,$\delta a_{j}^{\dagger}$]=[$%
\delta b$,$\delta b^{\dagger}$]=1, [$\delta a_{j}$,$H_{eff}$]$\thicksim$$%
\delta a_{j}$ and [$\delta b$,$H_{eff}$]$\thicksim$$\delta b$). We highlight
that the Eqs. (\ref{delta_b_dot}), (\ref{delta_cj_dot}) and (\ref{H-eff-1})
have been obtained by setting $a_{js}=-i\left\vert a_{js}\right\vert $ or
equivalently by taking the phase $\varphi _{j}$ of the $j^{th}$ input laser
field equal to $\varphi _{j}=-\arctan (2\Delta _{j}^{\prime }/\kappa _{j})$ 
\cite{Mauro(4)}. Now, we introduce the operators $\delta \tilde{b}$ and $%
\delta \tilde{a}_{j}$ defined by $\delta b=\delta \tilde{b}e^{-i\omega _{\mu
}t}$ and $\delta a_{j}=\delta \tilde{a}_{j}e^{i\Delta _{j}^{\prime }t}$ (for 
$\ j=1,2$) and using the Eqs. (\ref{delta_b_dot}) and (\ref{delta_cj_dot}),
one has 
\begin{eqnarray}
\ \partial _{t}\delta \tilde{b} &=&-\frac{\gamma _{\mu }}{2}\delta \tilde{b}%
+\sum_{j=1}^{2}(-1)^{j}\mathcal{G}_{j}\left( \delta \tilde{a}_{j}e^{i\left(
\omega _{\mu }+\Delta _{j}^{\prime }\right) t}-\delta \tilde{a}_{j}^{\dag
}e^{i\left( \omega _{\mu }-\Delta _{j}^{\prime }\right) t}\right) +\sqrt{%
\gamma _{\mu }}\tilde{b}^{in},  \label{delta_b_tilde_dot} \\
\partial _{t}\delta \tilde{a}_{j} &=&-\frac{\kappa _{j}}{2}\delta \tilde{a}%
_{j}+(-1)^{j+1}\mathcal{G}_{j}\left( \delta \tilde{b}e^{-i\left( \omega
_{\mu }+\Delta _{j}^{\prime }\right) t}+\delta \tilde{b}^{\dag }e^{i\left(
\omega _{\mu }-\Delta _{j}^{\prime }\right) t}\right) +\sqrt{\kappa _{j}}%
\tilde{a}_{j}^{in}\text{ \ for \ }j=1,2.  \label{delta_cj_tilde_dot}
\end{eqnarray}%
Next, we assume that the two cavities are driven at \textit{the red sideband}
($\Delta _{j}^{\prime }=-\omega _{\mu }$ for $j=1,2$) which corresponds to
the quantum states transfer \cite{Pinard,Tian-Clerk(3)-Sete} (we recall that
the $j^{th}$ laser detuning $\Delta _{j}$ has been defined as $\Delta
_{j}=\omega _{L_{j}}-\omega _{c_{j}}$). Further, in the resolved-sideband
regime, where the mechanical frequency $\omega _{\mu }$ of the movable
mirror is much larger than the $j^{th}$ cavity decay rate $\kappa _{j}$ ($%
\omega _{\mu }\gg \kappa _{1}$, $\kappa _{2}$), one can use the rotating
wave approximation (RWA) \cite{Marquardt,Clerk(2)}. Therefore, ignoring the
fast oscillating terms which rotating at the frequencies $\pm 2\omega _{\mu
} $ in Eqs. (\ref{delta_b_tilde_dot}) and (\ref{delta_cj_tilde_dot}), one
gets 
\begin{eqnarray}
\partial _{t}\delta \tilde{b} &=&-\frac{\gamma _{\mu }}{2}\delta \tilde{b}%
+\sum_{j=1}^{2}(-1)^{j}\mathcal{G}_{j}\delta \tilde{a}_{j}+\sqrt{\gamma
_{\mu }}\tilde{b}^{in},  \label{RWA_b_tilde_dot} \\
\partial _{t}\delta \tilde{a}_{j} &=&-\frac{\kappa _{j}}{2}\delta \tilde{a}%
_{j}+(-1)^{j+1}\mathcal{G}_{j}\delta \tilde{b}\ +\sqrt{\kappa _{j}}\tilde{a}%
_{j}^{in}\text{ \ for \ }j=1,2.  \label{RWA_cj_tilde_dot}
\end{eqnarray}%
We note that under the rotating wave approximation (RWA) with $\Delta
_{1,2}^{\prime }=-\omega _{\mu }$, Eq. (\ref{H-eff-1}) reduces to the
following simple expression 
\begin{equation}
H_{eff}^{\mathrm{RWA}}=i\sum_{j=1}^{2}(-1)^{j}\mathcal{G}_{j}\left[ \delta
a_{j}\delta b^{\dag }-\delta a_{j}^{\dag }\delta b\right],  \label{H-eff-2}
\end{equation}
from which one can derive the equations (\ref{RWA_b_tilde_dot}) and (\ref%
{RWA_cj_tilde_dot}).

\subsection{ Covariance matrix and Lyapunov equation}

In order to investigate the bipartite quantum correlations between the
different modes of the whole system, it is more convenient to transform Eqs.
(\ref{RWA_b_tilde_dot}) and (\ref{RWA_cj_tilde_dot}) in terms of the
quadratures operators of the three Gaussian modes (two optical modes and a
single mechanical mode) and their corresponding Hermitian input noise
operators. Thus, for the two optical cavities modes, we introduce 
\begin{eqnarray}
\delta \tilde{X}_{j} &=&(\delta \tilde{a}_{j}^{\dag }~+~\delta \tilde{a}%
_{j})/\sqrt{2}\text{ \ \ \ and \ \ \ \ }\delta \tilde{Y}_{j}=i(\delta \tilde{%
a}_{j}^{\dag }~-~\delta \tilde{a}_{j})/\sqrt{2}\text{ \ \ \ \ for \ }j=1,2,
\label{OQ} \\
\tilde{X}_{j}^{in} &=&(\tilde{a}_{j}^{in\dag }~+~\tilde{a}_{j}^{in})/\sqrt{2}%
\text{ \ \ \ and \ \ \ \ }\tilde{Y}_{j}^{in}=i(\tilde{a}_{j}^{in\dag }~-~%
\tilde{a}_{j}^{in})/\sqrt{2}\text{ \ \ \ \ \ \ for \ }j=1,2.
\label{Noise-OQ}
\end{eqnarray}%
In a similar way, we define for the single mechanical mode 
\begin{eqnarray}
\delta \tilde{q} &=&(\delta \tilde{b}^{\dag }~+~\delta \tilde{b})/\sqrt{2}%
\text{ \ \ \ \ and \ \ }\delta \tilde{p}=i(\delta \tilde{b}^{\dag }~-~\delta 
\tilde{b})/\sqrt{2},  \label{MQ} \\
\tilde{q}^{in} &=&(\tilde{b}^{in\dag }~+~\tilde{b}^{in})/\sqrt{2}\text{ \ \
\ and \ \ }\tilde{p}^{in}=i(\tilde{b}^{in\dag }~-~\tilde{b}^{in})/\sqrt{2}.
\label{NMQ}
\end{eqnarray}%
It is simple to check that the fluctuations of the quadratures operators
satisfy the following set of linear quantum Langevin equations%
\begin{eqnarray}
\partial _{t}\delta \tilde{X}_{j} &=&-\frac{\kappa _{j}}{2}\delta \tilde{X}%
_{j}+(-1)^{j+1}\mathcal{G}_{j}\delta \tilde{q}+\sqrt{\kappa _{j}}\tilde{X}%
_{j}^{in}\text{ \ \ \ for \ \ }j=1,2,  \label{delta--Xj-tilde-dot} \\
\partial _{t}\delta \tilde{Y}_{j} &=&-\frac{\kappa _{j}}{2}\delta \tilde{Y}%
_{j}+(-1)^{j+1}\mathcal{G}_{j}\ \delta \tilde{p}+\sqrt{\kappa _{j}}~\tilde{Y}%
_{j}^{in}\text{ \ for \ \ }j=1,2,\text{ }  \label{delta--Yj-tilde-dot} \\
\partial _{t}\delta \tilde{q} &=&\ \sum_{j=1}^{2}(-1)^{j}\mathcal{G}%
_{j}\delta \tilde{X}_{j}-\frac{\gamma _{\mu }}{2}\delta \tilde{q}+\sqrt{%
\gamma _{\mu }}\tilde{q}^{in},  \label{delta--q-tilde-dot} \\
\partial _{t}\delta \tilde{p} &=&\ \sum_{j=1}^{2}(-1)^{j}\mathcal{G}%
_{j}\delta \tilde{Y}_{j}-\frac{\gamma _{\mu }}{2}\text{\ }\delta \tilde{p}+%
\sqrt{\gamma _{\mu }}\tilde{p}^{in}.  \label{delta--p-tilde-dot}
\end{eqnarray}
Using the observables $\delta X_{j}$, $\delta Y_{j}$, $\delta q$ and $\delta
p$ defined by (\ref{OQ}) and (\ref{MQ}), the Hamiltonian (\ref{H-eff-2})
becomes 
\begin{equation}
\mathcal{H}_{eff}=\sum_{j=1}^{2}(-1)^{j}\mathcal{G}_{j}\left[ \delta
X_{j}\delta p-\delta Y_{j}\delta q\right],  \label{H-eff-QO}
\end{equation}
leading to linearized quantum Langevin equations (\ref{delta--Xj-tilde-dot}%
)-(\ref{delta--p-tilde-dot}). These equations can be written in the
following compact matrix form 
\begin{equation}
\partial _{t}u=Au+n,  \label{Compact form-ED}
\end{equation}%
with $u=(\delta \tilde{X}_{1},\delta \tilde{Y}_{1},\delta \tilde{X}%
_{2},\delta \tilde{Y}_{2},\delta \tilde{q},\delta \tilde{p})^{\mathrm{T}}$
and $n=(\sqrt{\kappa _{1}}\tilde{X}_{1}^{in},\sqrt{\kappa _{1}}\tilde{Y}%
_{1}^{in},\sqrt{\kappa _{2}}\tilde{X}_{2}^{in},\sqrt{\kappa _{2}}\tilde{Y}%
_{2}^{in},\sqrt{\gamma _{\mu }}\tilde{q}^{in},\sqrt{\gamma _{\mu }}\tilde{p}%
^{in})^{\mathrm{T}}$ are respectively the column vector of quadratures
fluctuations and the column vector of the noise sources. Moreover, the $%
6\times 6$ matrix $A$ in Eq. (\ref{Compact form-ED}) represents the drift
matrix of the system under investigation \cite{Mari(1)}. Introducing the $%
j^{th}$ multiphoton optomechanical cooperativity $\mathcal{C}_{j}$\ defined
as \cite{Marquardt,Regal} 
\begin{equation}
\mathcal{C}_{j}=\frac{4\mathcal{G}_{j}^{2}}{\kappa _{j}\gamma _{\mu }}=\frac{%
4g_{j}^{2}\bar{n}_{\mathrm{cav}}^{j}}{\kappa _{j}\gamma _{\mu }}=\frac{%
8\omega _{c_{j}}^{2}}{\gamma _{\mu }m_{\mu }\omega _{\mu }\omega
_{L_{j}}l_{j}^{2}}\frac{P_{j}}{\left[ \left( \frac{\kappa _{j}}{2}\right)
^{2}+\omega _{\mu }^{2}\right] }\text{ \ for \ \ }j=1,2\text{ ,}
\label{OCooperativity}
\end{equation}%
the drift matrix $A$ can be expressed as 
\begin{equation}
A=\frac{1}{2}\left( 
\begin{array}{cccccc}
-\kappa _{1} & 0 & 0 & 0 & \ \sqrt{\gamma _{\mu }\kappa _{1}\mathcal{C}_{1}}
& 0 \\ 
0 & -\kappa _{1} & 0 & 0 & 0 & \sqrt{\gamma _{\mu }\kappa _{1}\mathcal{C}_{1}%
} \\ 
0 & 0 & -\kappa _{2} & 0 & -\sqrt{\gamma _{\mu }\kappa _{2}\mathcal{C}_{2}}
& 0 \\ 
0 & 0 & 0 & -\kappa _{2} & 0 & -\sqrt{\gamma _{\mu }\kappa _{2}\mathcal{C}%
_{2}} \\ 
-\sqrt{\gamma _{\mu }\kappa _{1}\mathcal{C}_{1}} & 0 & \sqrt{\gamma _{\mu
}\kappa _{2}\mathcal{C}_{2}} & 0 & -\gamma _{\mu } & 0 \\ 
0 & -\sqrt{\gamma _{\mu }\kappa _{1}\mathcal{C}_{1}} & 0 & \sqrt{\gamma
_{\mu }\kappa _{2}\mathcal{C}_{2}} & 0 & -\gamma _{\mu }%
\end{array}%
\right) .  \label{drift}
\end{equation}%
We note that a weaker condition to reach the regime of strong optomechanical
coupling is given by $\mathcal{C}_{j}>>1$ (for $j=1,2$) \cite{Groblacker}.
The solution of Eq. (\ref{Compact form-ED}) can be written as \cite%
{G.Agarwal,Vitali(1)} 
\begin{equation}
u(t)=F(t)u(0)+\int_{0}^{t}dsF(s)n(t-s),  \label{u--Sol}
\end{equation}%
with $F(t)=\exp \left\{ At\right\} $. The system is stable and reaches its
steady state if and only when the real parts of all the eigenvalues of the
drift-matrix $A$ are negative, thus $F(\infty )=0$. The stability conditions
of the system can be obtained using the Routh--Hurwitz criterion \cite{Routh
and Hurwitz}. Due to the $6\times 6$ dimension of the drift matrix $A$ (see
Eq. (\ref{drift})), the explicit expressions of the stability conditions are
quite cumbersome and will not be reported here. We emphasize that all the
parameters chosen in this paper have been verified to satisfy the stability
conditions.

The quantum operators noises $a_{j}^{in}$ and $b^{in}$ are zero-mean quantum
Gaussian noises and the dynamics has been linearized (see Eqs. [(\ref%
{delta--Xj-tilde-dot})-(\ref{delta--p-tilde-dot})]). So, the steady state of
the system is a zero-mean tripartite Gaussian state with zero mean
fluctuations \cite{Vitali(2)}. The system is completely specified by its $%
6\times 6$ covariance matrix (CM) $V$ \ \cite{Olivares(2)}, with matrix
elements 
\begin{equation}
V_{ii^{\prime }}=(\langle u_{i}(\infty )u_{i^{\prime }}(\infty
)+u_{i^{\prime }}(\infty )u_{i}(\infty )\rangle )/2,  \label{CM-elements0}
\end{equation}%
where $u=\bigg(\delta \tilde{X}_{1}(\infty ),\delta \tilde{Y}_{1}(\infty
),\delta \tilde{X}_{2}(\infty ),\delta \tilde{Y}_{2}(\infty ),\delta \tilde{q%
}(\infty ),\delta \tilde{p}(\infty )\bigg)^{\mathrm{T}}$ is the vector of
continuous variable fluctuation operators in the steady state $(t\rightarrow
\infty )$. We note that $V$ is a real, symmetric, positive definite matrix 
\cite{Adesso(3)}. When the system is stable and using Eq. (\ref{u--Sol}),
the covariance matrix elements write%
\begin{equation}
V_{ii^{\prime }}=\sum_{k,k^{\prime }}\int_{0}^{\infty }ds\int_{0}^{\infty
}ds^{\prime }F_{^{ik}}(s)F_{^{i^{\prime }k^{\prime }}}(s^{\prime })\Phi
_{^{kk^{\prime }}}(s-s^{\prime }),  \label{CM-elements}
\end{equation}%
where $\Phi _{^{kk^{\prime }}}(s-s^{\prime })=(\langle n_{k}(s)n_{k^{\prime
}}(s^{\prime })+n_{k^{\prime }}(s^{\prime })n_{k}(s)\rangle
)/2=D_{^{kk^{\prime }}}\delta (s-s^{\prime })$ are the components of the
diffusion matrix $D$ of the stationary noise correlation functions \cite%
{Mari(1)}. Using the correlation properties of the noise operators given by
the Eqs. [(\ref{ther_noise_1})-(\ref{S_4})], we obtain 
\begin{equation}
D=\left( 
\begin{array}{cccccc}
\kappa _{1}\left( N+\frac{1}{2}\right) \text{\ } & \ 0 & \sqrt{\kappa
_{1}\kappa _{2}}M & \ 0 & \ 0 & \ 0 \\ 
\ 0 & \kappa _{1}\left( N+\frac{1}{2}\right) & 0\  & -\sqrt{\kappa
_{1}\kappa _{2}}M & \ 0 & \ 0 \\ 
\sqrt{\kappa _{1}\kappa _{2}}M & 0\  & \kappa _{2}\left( N+\frac{1}{2}\right)
& \ 0 & \ 0 & \ 0 \\ 
\ 0 & -\sqrt{\kappa _{1}\kappa _{2}}M & \ 0 & \kappa _{2}\left( N+\frac{1}{2}%
\right) \text{\ } & \ 0 & \ 0 \\ 
\ 0 & \ 0 & \ 0 & \ 0 & \gamma _{\mu }\left( n_{\mathrm{th}}+\frac{1}{2}%
\right) \text{\ } & \ 0 \\ 
\ 0 & \ 0 & \ 0 & \ 0 & \ 0 & \gamma _{\mu }\left( n_{\mathrm{th}}+\frac{1}{2%
}\right)%
\end{array}%
\right) .  \label{Diffusion}
\end{equation}%
From Eq. (\ref{CM-elements}), the covariance matrix $V$ writes also as%
\begin{equation}
V=\int_{0}^{\infty }dsF(s)DF(s)^{\mathrm{T}}.  \label{CM-intgr-form}
\end{equation}
When the system is stable ($F(\infty )=0$), Eq. (\ref{CM-intgr-form}) is
equivalent to the Lyapunov equation for the steady-state (CM) \cite{Mari(2)} 
\begin{equation}
AV+VA^{\mathrm{T}}=-D.  \label{Lyapunov}
\end{equation}
It is clear that the Eq. (\ref{Lyapunov}) is linear for $V$, thus it can be
straightforwardly solved, but the explicit expression of $V$ is too
cumbersome and can not be reported here.

\section{Bipartite and tripartite optomechanical entanglement\label{sec3}}

\subsection{The Simon criterion as witness of entanglement}

\noindent In the systems of continuous-variables (CVs), the investigation of
the entanglement properties has been the object of study in a number of
recent publications in bipartite systems \cite%
{Mauro(1),Eisert(2)-Tombisi(1),Pinard,G.Agarwal,Vitali(1),Grebogi-Eisert(3),Daoud,Giovanitti(1)}%
. The quantum correlations in a tripartite optical system were reported in 
\cite{M.G.Paris(3)} and more recently many other proposals focused
particularly into the field of optomechanics \cite%
{Mauro(3),Mauro(5),Vitali(3)-Vitali(4),Aggarwal-Mabdi-Naderi-Jie
Song-E.Wu,Mazzola(2)}. We notice that an interesting review of the theory of
(CVs) entanglement was concisely given in \cite{Eisert(4)}. The covariance
matrix $V$ which is solution of the Lyapunov equation (Eq. (\ref{Lyapunov}%
)), can be written in the $3\times 3$ block following form 
\begin{equation}
V=\left[ V_{ii^{\prime }}\right] _{6\times 6}=\left( 
\begin{array}{ccc}
B_{o_{1}} & C_{o_{1}o_{2}} & C_{o_{1}m} \\ 
C_{o_{1}o_{2}}^{\mathrm{T}} & B_{o_{2}} & C_{o_{2}m} \\ 
C_{o_{1}m}^{\mathrm{T}} & C_{o_{2}m}^{\mathrm{T}} & B_{m}%
\end{array}%
\right) ,  \label{Global CM}
\end{equation}%
where $B_{_{j^{\prime }}}$ is a $2\times 2$ matrix that describes the local
properties of the $j^{\prime }$-mode. Whereas, $C_{_{j^{\prime }j^{\prime
\prime }}}(j^{\prime }\neq j^{\prime \prime }=o_{1},o_{2},m)$ describes the
intermode correlations. Thus, the reduced covariance matrix describing the
correlations between $j^{\prime }$ and $j^{\prime \prime }$ modes is given
by 
\begin{equation}
\left[ V_{j^{\prime }j^{\prime \prime }}\right] _{4\times 4}=\left( 
\begin{array}{cc}
B_{j^{\prime }} & C_{_{j^{\prime }j^{\prime \prime }}} \\ 
C_{_{j^{\prime }j^{\prime \prime }}}^{\mathrm{T}} & B_{j^{\prime \prime }}%
\end{array}%
\right) .  \label{Sub-CM}
\end{equation}%
Due to the Gaussian nature of the system under investigation, the bipartite
entanglement between the two modes $j^{\prime }$ and $j^{\prime \prime }$ $%
(j^{\prime }\neq j^{\prime \prime }=o_{1},o_{2},m)$ can be quantified in
terms of the Simon's necessary and sufficient entanglement nonpositive
partial transpose criterion of the Gaussian states \cite{Simon}. According
to this end, the two modes $j^{\prime }$ and $j^{\prime \prime }$ $%
(j^{\prime }\neq j^{\prime \prime }=o_{1},o_{2},m)$ are entangled if and
only if $\eta _{_{j^{\prime }j^{\prime \prime }}}^{-}<1/2$. The smallest
symplectic eigenvalue $\eta _{_{j^{\prime }j^{\prime \prime }}}^{-}$ which
obtained by the partial transpose of the $4\times 4$ covariance matrix Eq. (%
\ref{Sub-CM}) is given by \cite{Vidal,Adesso(1),Eisert(5)} 
\begin{equation}
\eta _{_{j^{\prime }j^{\prime \prime }}}^{-}=\sqrt{\frac{\Delta _{j^{\prime
}j^{\prime \prime }}-\sqrt{\Delta _{_{j^{\prime }j^{\prime \prime
}}}^{2}-4\det V_{_{j^{\prime }j^{\prime \prime }}}}}{2}},  \label{S-S-EGV}
\end{equation}%
with $\Delta _{_{j^{\prime }j^{\prime \prime }}}=\det B_{j^{\prime }}+\det
B_{j^{\prime \prime }}-2\det C_{_{j^{\prime }j^{\prime \prime }}}$, where
the three $2\times 2$ submatrices $B_{j^{\prime }},$ $B_{j^{\prime \prime }}$
and $C_{_{j^{\prime }j^{\prime \prime }}}$ can be extracted from Eq. (\ref%
{Sub-CM}).

\subsection{Stationary bipartite and tripartite entanglement analysis}

\textbf{\textit{3.2.1~~Entanglement analysis versus the thermal effect}}%
\newline

We now analyze the stationary entanglement distribution among the three
possible bipartite subsystems using the smallest symplectic eigenvalues $%
\eta _{_{j^{\prime }j^{\prime \prime }}}^{-}$ ($j^{\prime }\neq j^{\prime
\prime }=o_{1},o_{2},m$) given by (\ref{S-S-EGV}). In what follows, $\eta
_{_{mo_{1}}}^{-}$, $\eta _{_{mo_{2}}}^{-}$ and $\eta _{_{o_{1}o_{2}}}^{-}$
denote respectively the smallest symplectic eigenvalue witnesses the
entanglement $m-o_{1}$ between the mechanical mode $m$ and the optical mode $%
o_{1}$, the entanglement $m-o_{2}$ between the mechanical mode $m$ and the
optical mode $o_{2}$ and finally, the entanglement $o_{1}-o_{2}$ between the
two optical cavities modes $o_{1}$ and $o_{2}$. The behavior of the three
bipartite entanglements have been analyzed under influence of the
temperature $T$ of the thermal bath of the movable mirror, the squeezing
parameter $r$ and also the optomechanical cooperativity $\mathcal{C}_{1}$ of
the right cavity. For simplicity, we take all the two cavities parameters to
be identical, except the $j^{th}$ input power laser $P_{j}$, where we have
fixed $P_{2}=2P_{1}$ (without loss of generality) or equivalently $\mathcal{C%
}_{2}=2\mathcal{C}_{1}$ (see Eq. (\ref{OCooperativity})). We start with the
influence of the temperature on the three entanglements. In order to do
realistic estimation, we used parameters from recent optomechanical
experiment \cite{Groblacker}: 
\begin{figure}[tbh]
\textrm{\centering
\begin{minipage}[htb]{2in}
\centering
\includegraphics[width=2in]{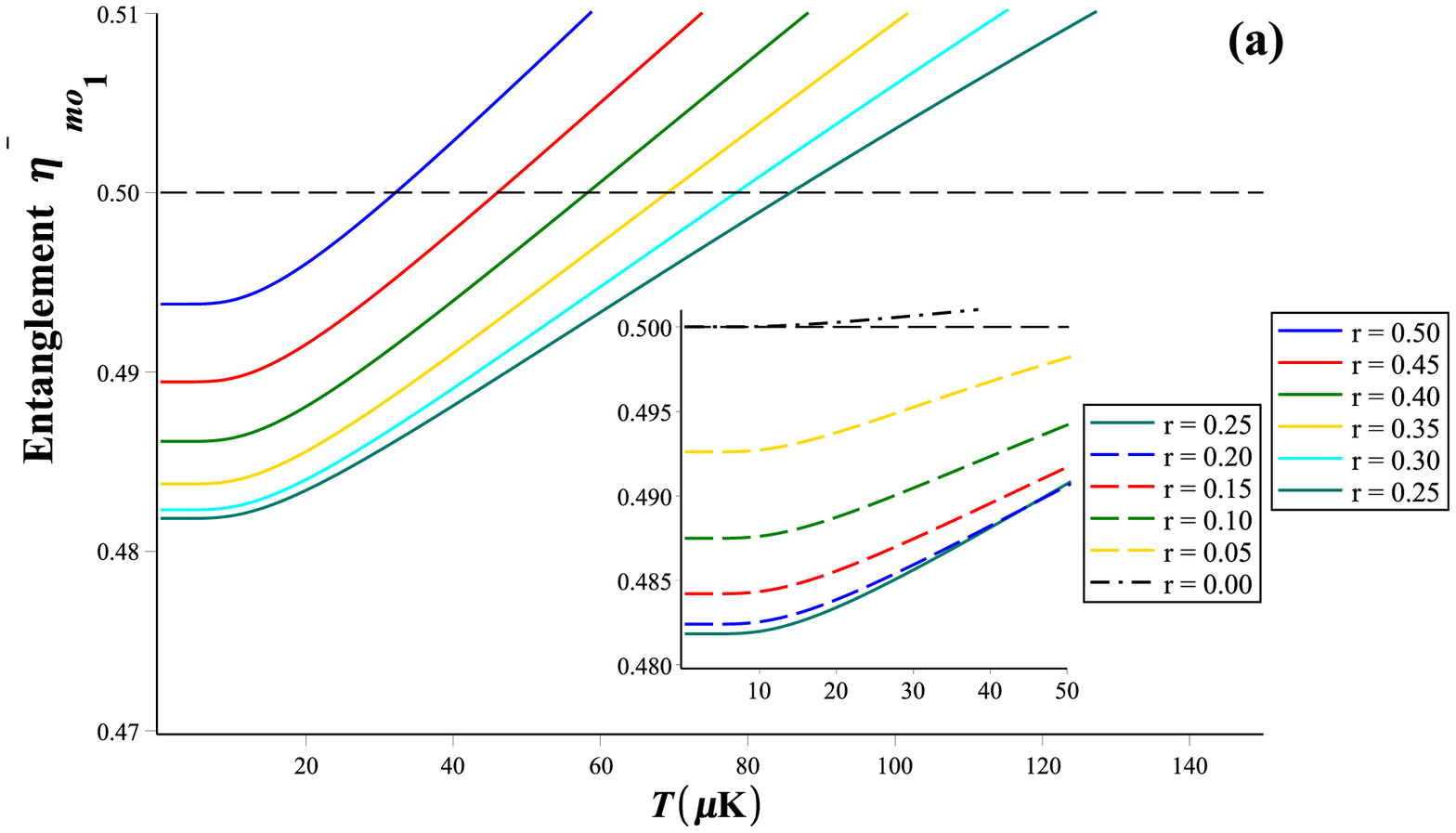}
\end{minipage}
\begin{minipage}[htb]{2in}
\centering
 \includegraphics[width=2in]{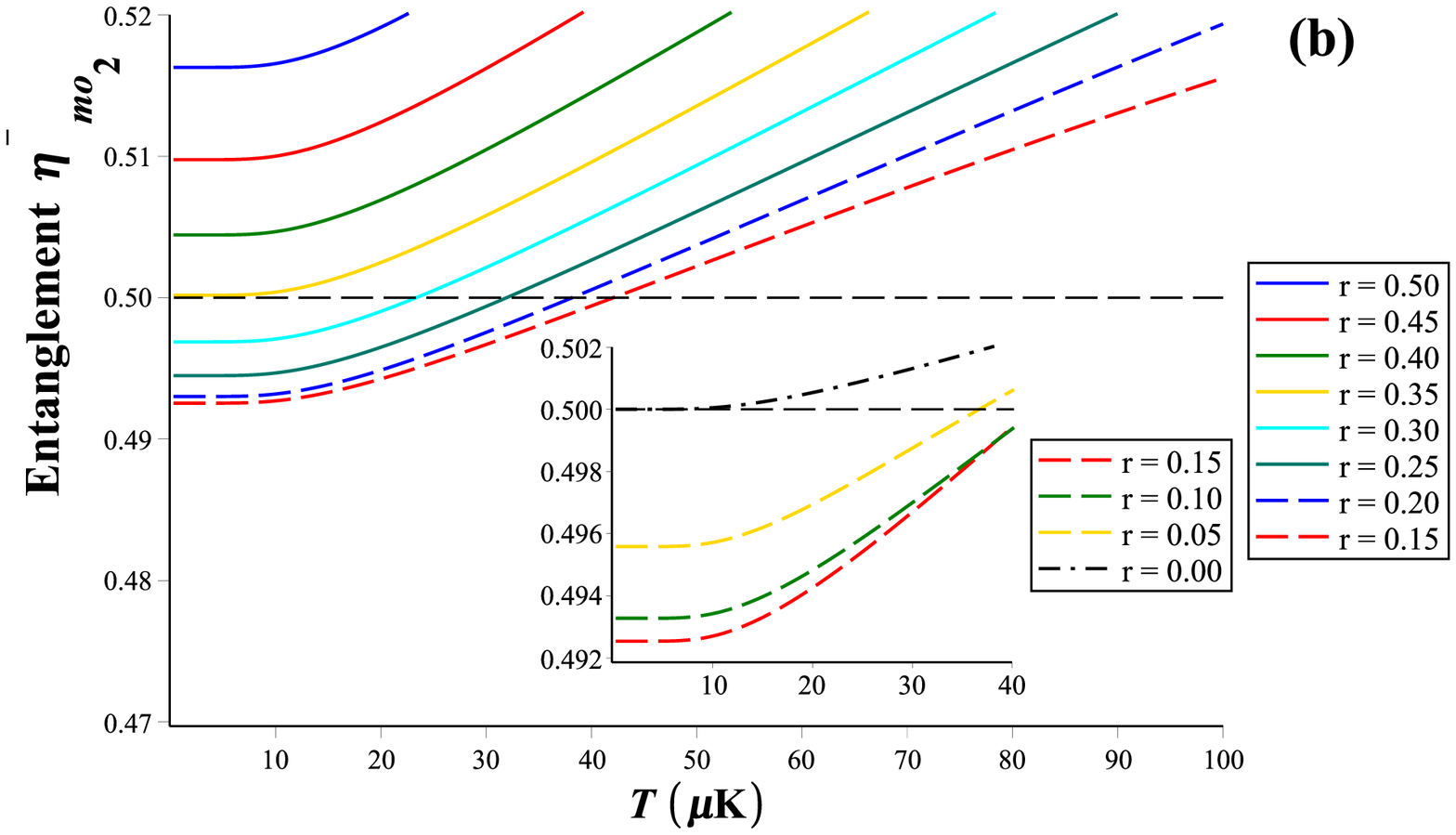}
\end{minipage}
\begin{minipage}[htb]{2in}
\centering
 \includegraphics[width=2in]{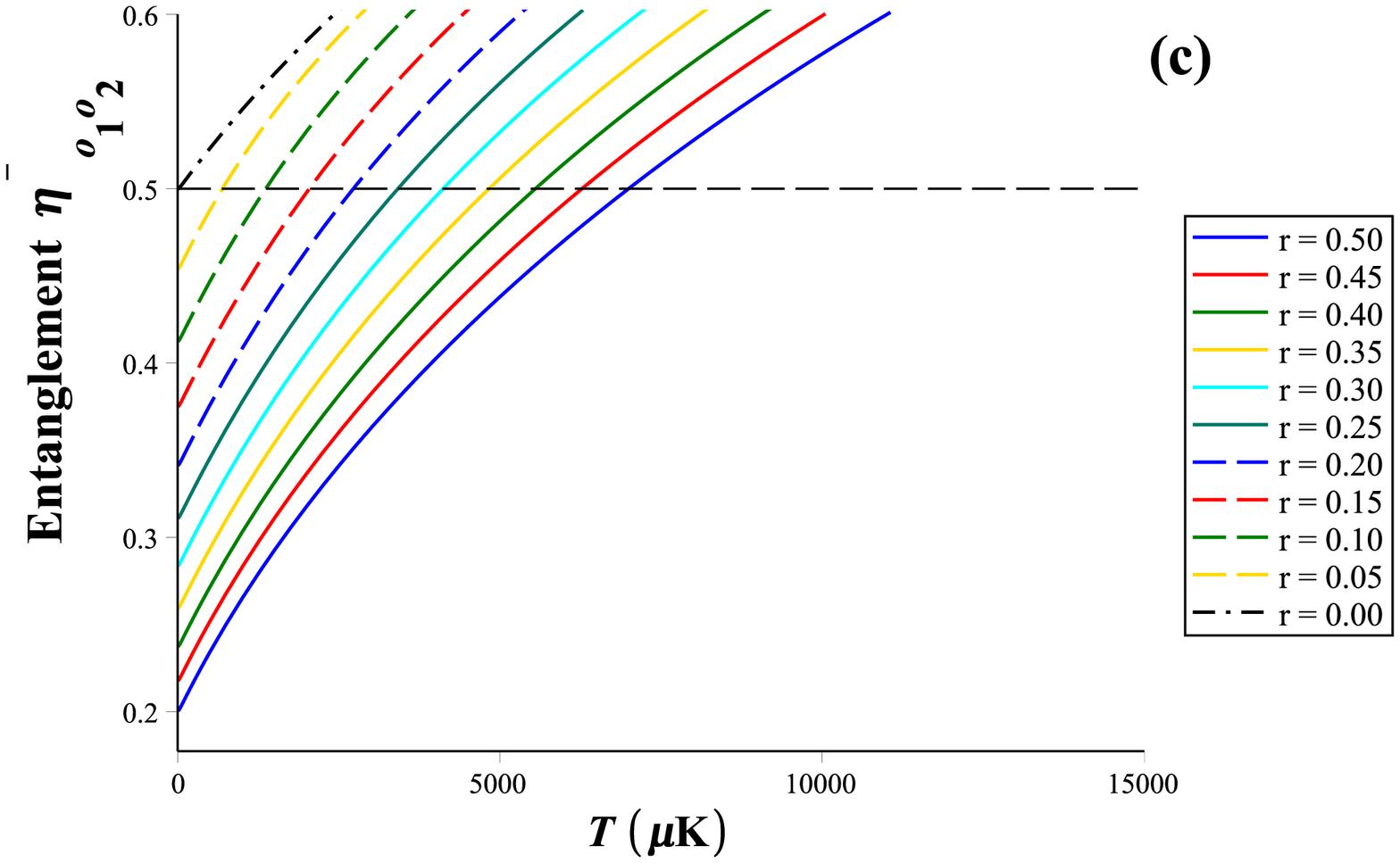}
\end{minipage}
}
\caption{ Mechanical bath temperature dependence of the smallest symplectic
eigenvalue $\protect\eta _{j^{\prime }j^{\prime \prime }}^{-}$ used as a
witness of the bipartite entanglement between, (a): the mechanical mode $m$
and the optical mode $o_{1}$, (b): the mechanical mode $m$ and the optical
mode $o_{2}$, (c): the two optical modes $o_{1}$ and $o_{2}$. All the two
cavities parameters are identical, except the $j^{th}$ input power laser $%
P_{j}$, where we have taken $P_{2}=2P_{1}$ or equivalently $\mathcal{C}_{2}=2%
\mathcal{C}_{1}$. In the panels (a), (b) and (c), each curve corresponds to
a given value of the squeezing parameter $r$. The horizontal dashed line ($%
\protect\eta _{j^{\prime }j^{\prime \prime }}^{-}=0.5$) represents the
boundary between the entangled states $\protect\eta _{j^{\prime }j^{\prime
\prime }}^{-}<0.5$ and the others separable $\protect\eta _{j^{\prime
}j^{\prime \prime }}^{-}>0.5$. Unlike the two hybrid subsystems, significant
entanglement is found over a wide range of temperatures in the purely
optical subsystem (up to $T=6.5~m\mathrm{K}$ for $r=0.5$ in panel (c)). The
two insets of panels (a) and (b) reveal the effect of low squeezing $r$ on
the $m-o_{1}$ and $m-o_{2}$ entanglements. We emphasize that, the energies
decay rates $\protect\kappa _{1,2}$, the mechanical damping rate $\protect%
\gamma _{\protect\mu }$ and the optomechanical cooperativity $\mathcal{C}%
_{1} $ are fixed respectively as $\protect\kappa _{1,2}=2\protect\pi \times
215\times 10^{3}~\mathrm{Hz}$, $\protect\gamma _{\protect\mu }=2\protect\pi %
.1500~\mathrm{Hz}$ and $\mathcal{C}_{1}=35$. }
\label{Fig.2}
\end{figure}
the wavelength of the lasers $\lambda _{1,2}=1064~\mathrm{nm}$ so\ $\omega
_{L_{1,2}}=2\pi \times 2.82\times 10^{14}~\mathrm{Hz}$, $P_{1}=10~\mathrm{mW}
$ ($P_{2}=2P_{1}=20~\mathrm{mW)}$, $l_{1,2}=25~\mathrm{mm}$, $\kappa
_{1,2}=2\pi \times 215\times 10^{3}~\mathrm{Hz}$, $\omega _{\mu }=2\pi
\times 947\times 10^{3}~\mathrm{Hz}$, $m_{\mu }=145~\mathrm{ng}$ and $\omega
_{c_{1,2}}\approx $ $3.\,\allowbreak 5\times 10^{15}$ $\mathrm{Hz}$. For the
mechanical damping rate $\gamma _{\mu }$, we have used $\gamma _{\mu
}\approx 2\pi \times 1.5\times 10^{3}~\mathrm{Hz}$, which is very comparable
to the value that used in \cite{Murch}. Next, using the explicit expression
of the $j^{th}$ optomechanical cooperativity $\mathcal{C}_{j}$ given by Eq. (%
\ref{OCooperativity}), one has $\mathcal{C}_{1}$ $\approx $ $35$ so $%
\mathcal{C}_{2}$ $=2\mathcal{C}_{1}\approx 70$. The dependence of the three
bipartite entanglements $m-o_{1}$, $m-o_{2}$ and $o_{1}-o_{2}$ on the
mechanical bath temperature for various squeezing $r$ is presented in Fig. %
\ref{Fig.2}. For a given squeeze $r$, as the environmental temperature
increases, the amount of the $j^{\prime }-j^{\prime \prime }$ \textit{%
entanglement} monotonically decreases ($\eta _{_{j^{\prime }j^{\prime \prime
}}}^{-}$ increases) due to the thermal fluctuations. Consequently, above a
critical temperature $T_{c}$, the $j^{\prime }-j^{\prime \prime }$
entanglement disappears completely as expected ($T_{c}$ defined as: $T>T_{c}$%
, $\eta _{_{j^{\prime }j^{\prime \prime }}}^{-}>1/2$ for a given squeeze $r$%
). Such a behavior is commonly known as \textit{entanglement sudden death}
(ESD) \cite{MD-Qasimi}. Comparing with the $m-o_{1}$ and $m-o_{2}$
entanglements, it can be clearly seen from Fig. \ref{Fig.2} that the $%
o_{1}-o_{2}$ entanglement is considerably large and more robust against the
thermal noises enhanced by high temperatures and enough squeezing.
Obviously, the panels (a), (b) and (c) in Fig. \ref{Fig.2}, show that for
zero squeezing ($r=0$), $\eta _{_{j^{\prime }j^{\prime \prime }}}^{-}$ is
always upper than 1/2 whatever $T$ and regardless $j^{\prime }$ and $%
j^{\prime \prime }$ ($j^{\prime }\neq j^{\prime \prime }=o_{1},o_{2},m$),
meaning that no entanglement occurs in any bipartite subsystem. Whereas, the
two modes $j^{\prime }$ and $j^{\prime \prime }$ become entangled if we
inject the squeezed light, indicating quantum fluctuations transfer from the
two-mode squeezed light to the three subsystems. In addition, the two insets
of panels (a) and (b) reveal that low values of the squeezing $r$ enhance
the entanglements $m-o_{1}$ and $m-o_{2}$. In contrast, high values of the
squeezing $r$ induce the entanglement degradation in the two hybrid
subsystems. Finally, it is interesting to remark that for small values of $T$
and $r$ ($T<20{\mu }\mathrm{K}$, $0<r<0.3$), the three-bipartite
entanglements $m-o_{1}$, $m-o_{2}$ and $o_{1}-o_{2}$ can be observed
simultaneously, which confirms the existence of strong correlations
distribution between the three optomechanical modes. \newline

\textbf{\textit{3.2.2~~Entanglement analysis versus the squeezing effect}} 
\newline

\begin{figure}[tbh]
\textrm{\textrm{\centering
\begin{minipage}[htb]{2in}
\centering
\includegraphics[width=2in]{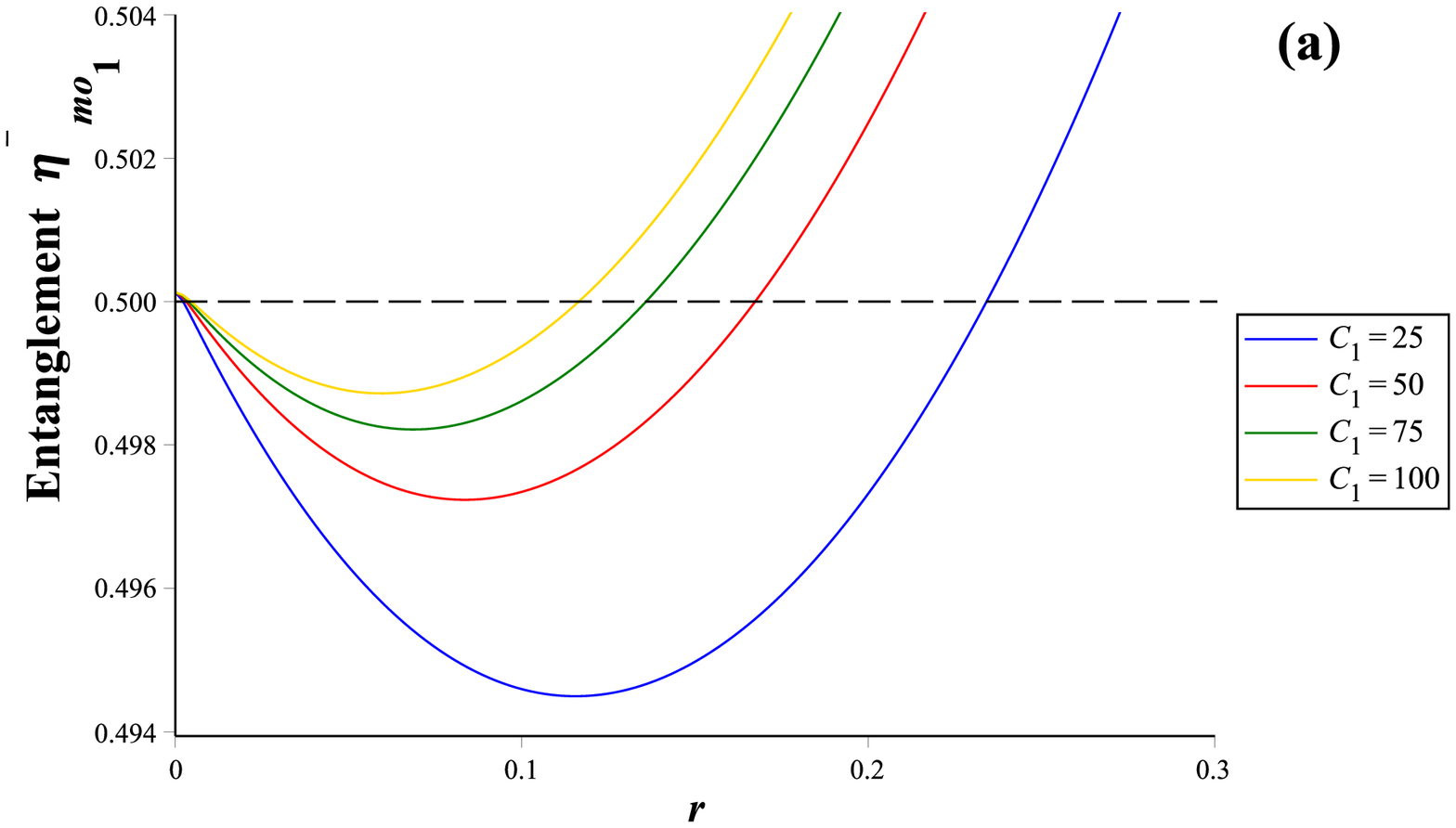}
\end{minipage}
\begin{minipage}[htb]{2in}
\centering
 \includegraphics[width=2in]{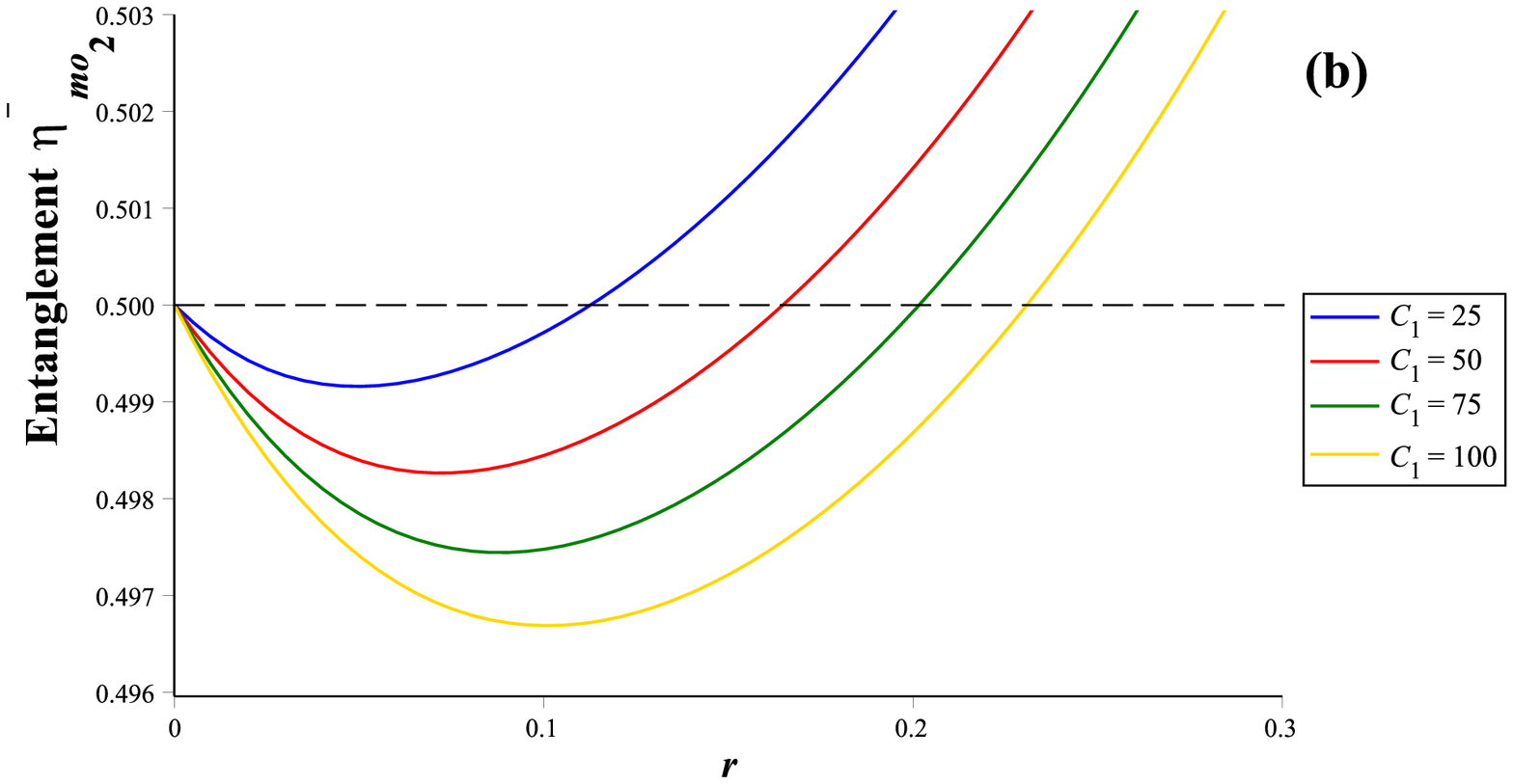}
\end{minipage}
\begin{minipage}[htb]{2in}
\centering
 \includegraphics[width=2in]{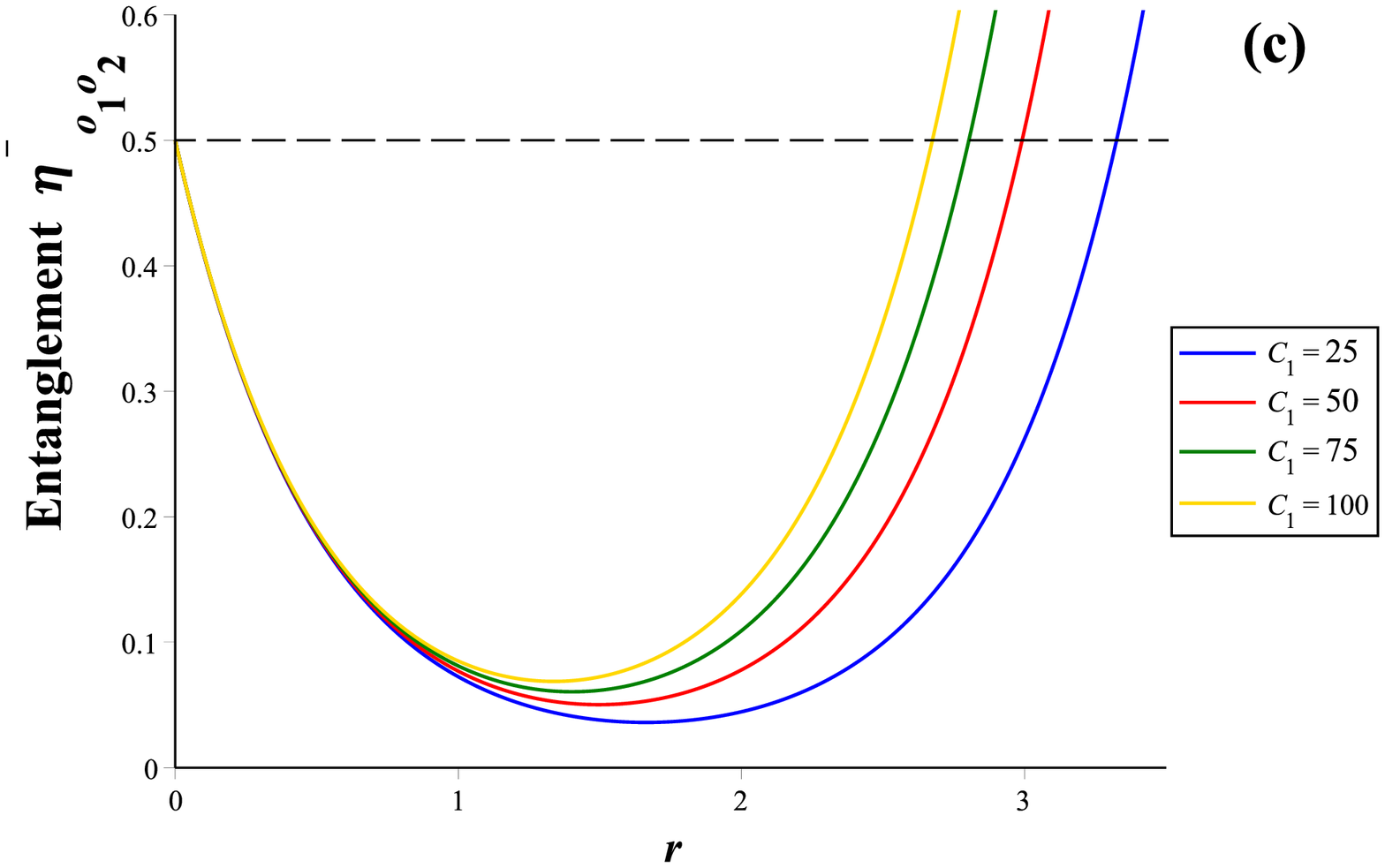}
\end{minipage}
} }
\caption{The three bipartite entanglements versus the squeezing parameter $r$
for various values of the dimensionless optomechanical cooperativity $%
\mathcal{C}_{1}$ of the right cavity. As in Fig. \protect\ref{Fig.2}, we
have taken $\mathcal{C}_{2}=2\mathcal{C}_{1}$ and $\protect\kappa _{1}=%
\protect\kappa _{2}=$ $2\protect\pi \times 215\times 10^{3}~\mathrm{Hz}$.
The panels (a), (b) and (c) correspond respectively to the bipartite
entanglement between : (a) the mechanical mode $m$ and the optical mode $%
o_{1}$ ($\protect\eta _{mo_{1}}^{-} $), (b) the mechanical mode $m$ and the
optical mode $o_{2}$ ($\protect\eta _{mo_{2}}^{-}$) and finally, (c) the two
optical modes $o_{1}$ and $o_{2}$ ($\protect\eta _{o_{1}o_{2}}^{-}$). Here
we used $\protect\gamma _{\protect\mu }=2\protect\pi .140~\mathrm{Hz}$ as a
value of the mechanical damping rate \protect\cite{Groblacker}. For the mean
thermal photons number $n_{\mathrm{th}} $, we have used $n_{\mathrm{th}%
}=10^{-3}$ ($T\simeq 6.\,\allowbreak 5~\protect\mu \mathrm{K}$). In the
three panels (a), (b) and (c), the horizontal dashed line corresponds to $%
\protect\eta _{_{j^{\prime }j^{\prime \prime }}}^{-}=1/2$ below which the
two modes ($j^{\prime }-j^{\prime \prime } $) are entangled. This figure
shows that the three entanglements $m-o_{1}$, $m-o_{2}$ and $o_{1}-o_{2}$
have the resonance-like behavior with respect to the squeezing values $r$.
Large entanglement has been detected up to $r\simeq 3.2$ in the purely
optical subsystem for $\mathcal{C}_{1}=25$. }
\label{Fig.3}
\end{figure}
The squeezed light effect on the three bipartite entanglements $m-o_{1}$, $%
m-o_{2}$ and $o_{1}-o_{2}$ quantified respectively by $\eta _{_{mo_{1}}}^{-}$%
, $\eta _{_{mo_{2}}}^{-}$ and $\eta _{_{o_{1}o_{2}}}^{-}$ is presented in
Fig. \ref{Fig.3}. For the mechanical damping rate $\gamma _{\mu }$ and the $%
j^{th}$ energy decay rate $\kappa _{j}$, we have used respectively $\gamma
_{\mu }=2\pi \times 140~\mathrm{Hz}$ and $\kappa _{1}=\kappa _{2}=$ $2\pi
\times 215\times 10^{3}~\mathrm{Hz}$ \cite{Groblacker}. For the mean thermal
photons number $n_{\mathrm{th}}$ we used $n_{\mathrm{th}}=10^{-3}$ or
equivalently $T\simeq 6.\,\allowbreak 5~\mu \mathrm{K}$. In the panels (a),
(b) and (c), each curve corresponds to a given value of the optomechanical
cooperativity $\mathcal{C}_{1}$ ($\mathcal{C}_{2}$ $=2\mathcal{C}_{1}$)$.$
As depicted in the panels (a), (b) and (c), there would be no entanglement
in any subsystem if $r=0$. In addition, Fig. \ref{Fig.3} reveals that the
three bipartite entanglements $m-o_{1}$, $m-o_{2}$ and $o_{1}-o_{2}$ have
the resonance-like behavior with respect to the squeezing parameter $r$ for
a fixed value of $\mathcal{C}_{1}$. Indeed, by increasing the squeezing
parameter $r$, the three functions $\eta _{_{mo_{1}}}^{-}$, $\eta
_{_{mo_{2}}}^{-}$ and $\eta _{_{o_{1}o_{2}}}^{-}$ decrease gradually (the
entanglements $m-o_{1}$, $m-o_{2}$ and $o_{1}-o_{2}$ increase) reaching
their minimum for a specific value of $r$ denoted $r_{0}$ ($r_{0}$ depends
both on the subsystem class and the fixed value of $\mathcal{C}_{1}$).
Furthermore, for $r>$ $r_{0}$ the functions $\eta _{_{mo_{1}}}^{-}$, $\eta
_{_{mo_{2}}}^{-}$ and $\eta _{_{o_{1}o_{2}}}^{-}$ increase with $r$ (the
entanglements $m-o_{1}$, $m-o_{2}$ and $o_{1}-o_{2}$ decrease) and quickly
become up to $1/2$, which corresponds to the entanglements degradation. The
resonance-like behavior can be explicated as follows: for $0<r<r_{0}$, the
photons number in the two cavities increases, which enhances the
optomechanical coupling by means of radiation pressure and consequently
leads to robust entanglement \cite{G.Agarwal}. In contrast, for $r>$ $r_{0}$
the input thermal noise affecting each cavity becomes important and more
aggressive causing the entanglement degradation \cite{Mauro(1)}. More
important, Fig. \ref{Fig.3} shows that among the three bipartite
entanglements $m-o_{1}$, $m-o_{2}$ and $o_{1}-o_{2}$, the largest and robust
one against the thermal noise enhanced by broadband squeezed light is also
that between the two optical cavities modes, even though they are uncoupled.%
\newline

\textbf{\textit{3.2.3~~Entanglement analysis versus the optomechanical
coupling}}\newline

\begin{figure}[tbh]
\textrm{\textrm{\centering
\begin{minipage}[htb]{2in}
\centering
\includegraphics[width=2in]{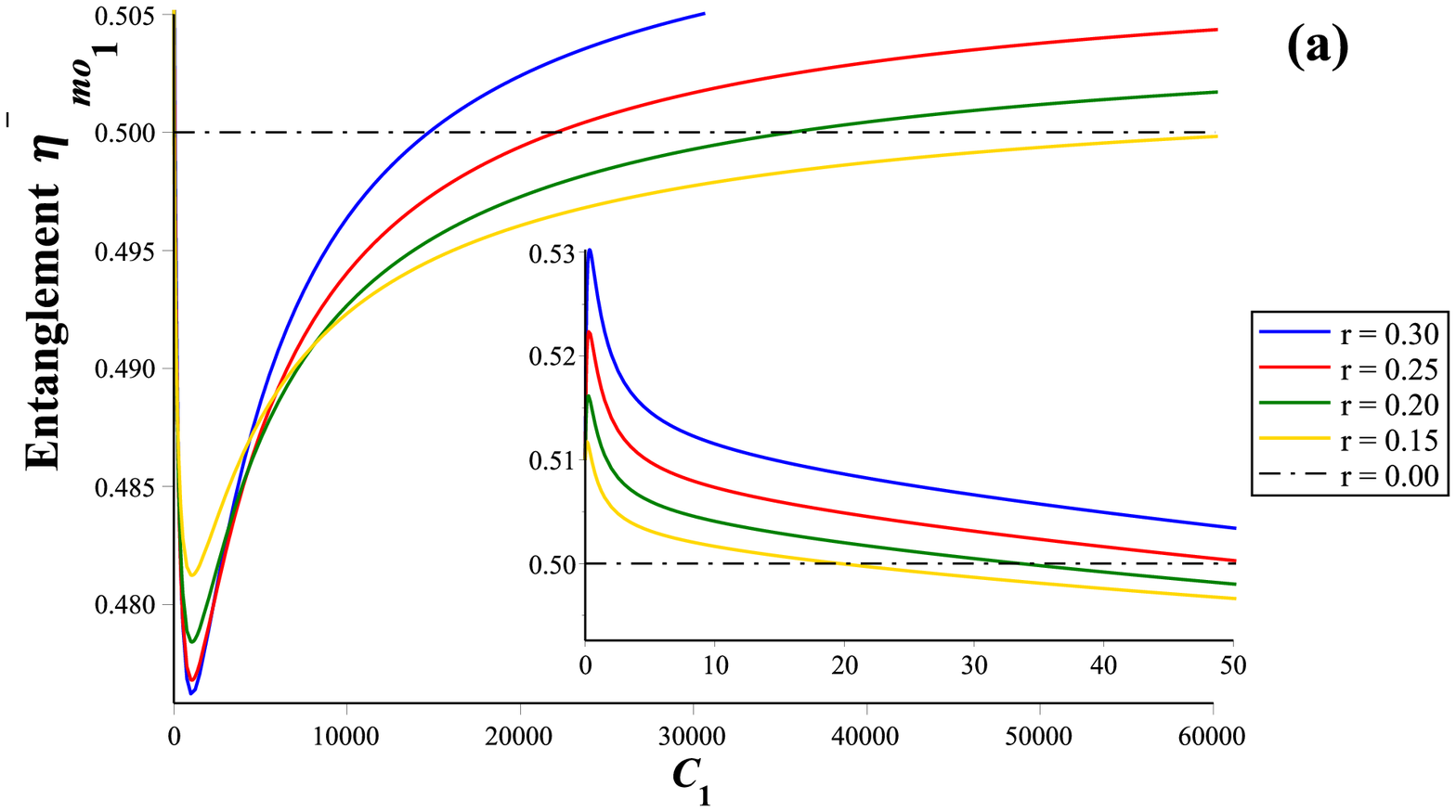}
\end{minipage}
\begin{minipage}[htb]{2in}
\centering
 \includegraphics[width=2in]{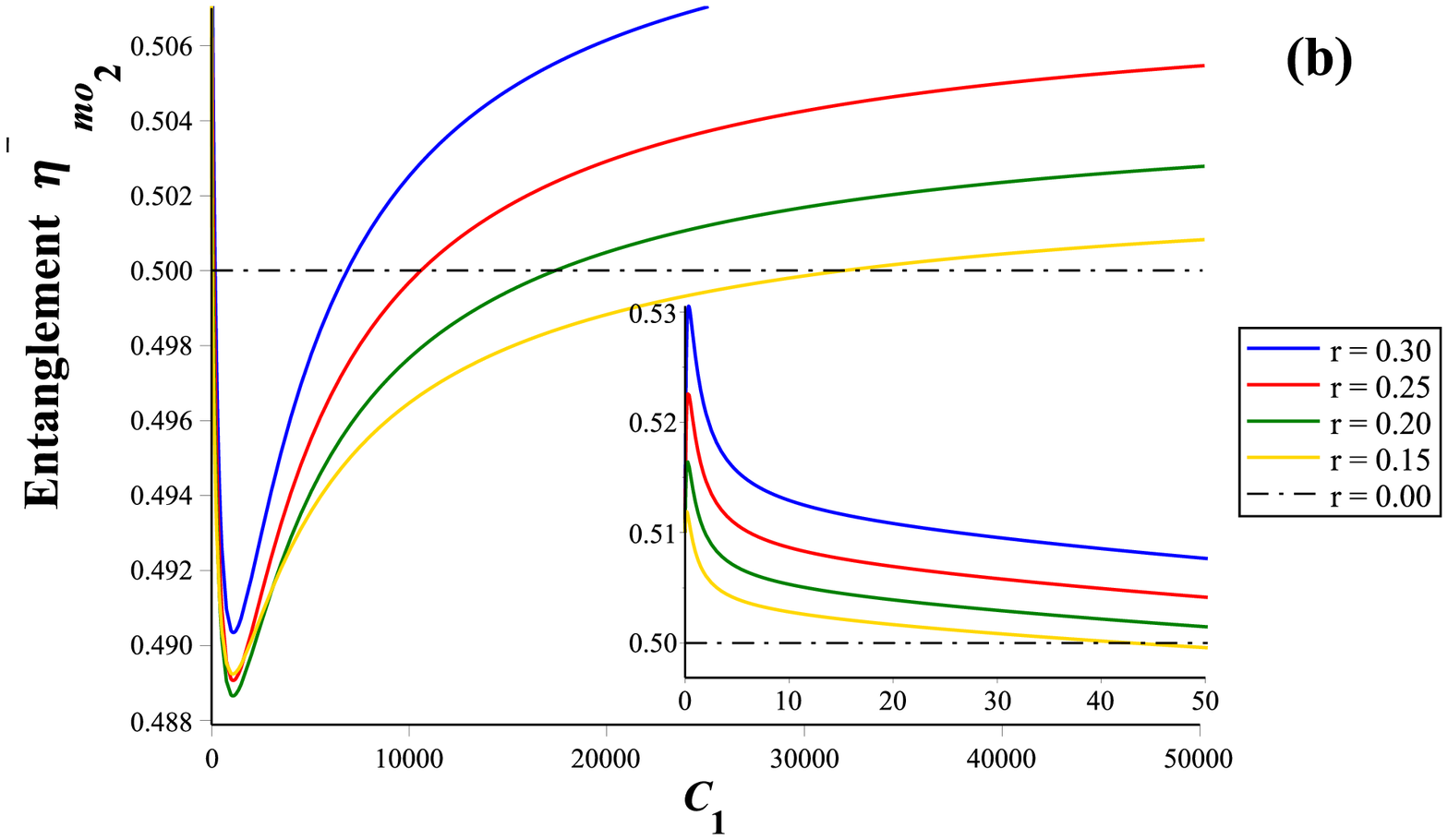}
\end{minipage}
\begin{minipage}[htb]{2in}
\centering
 \includegraphics[width=2in]{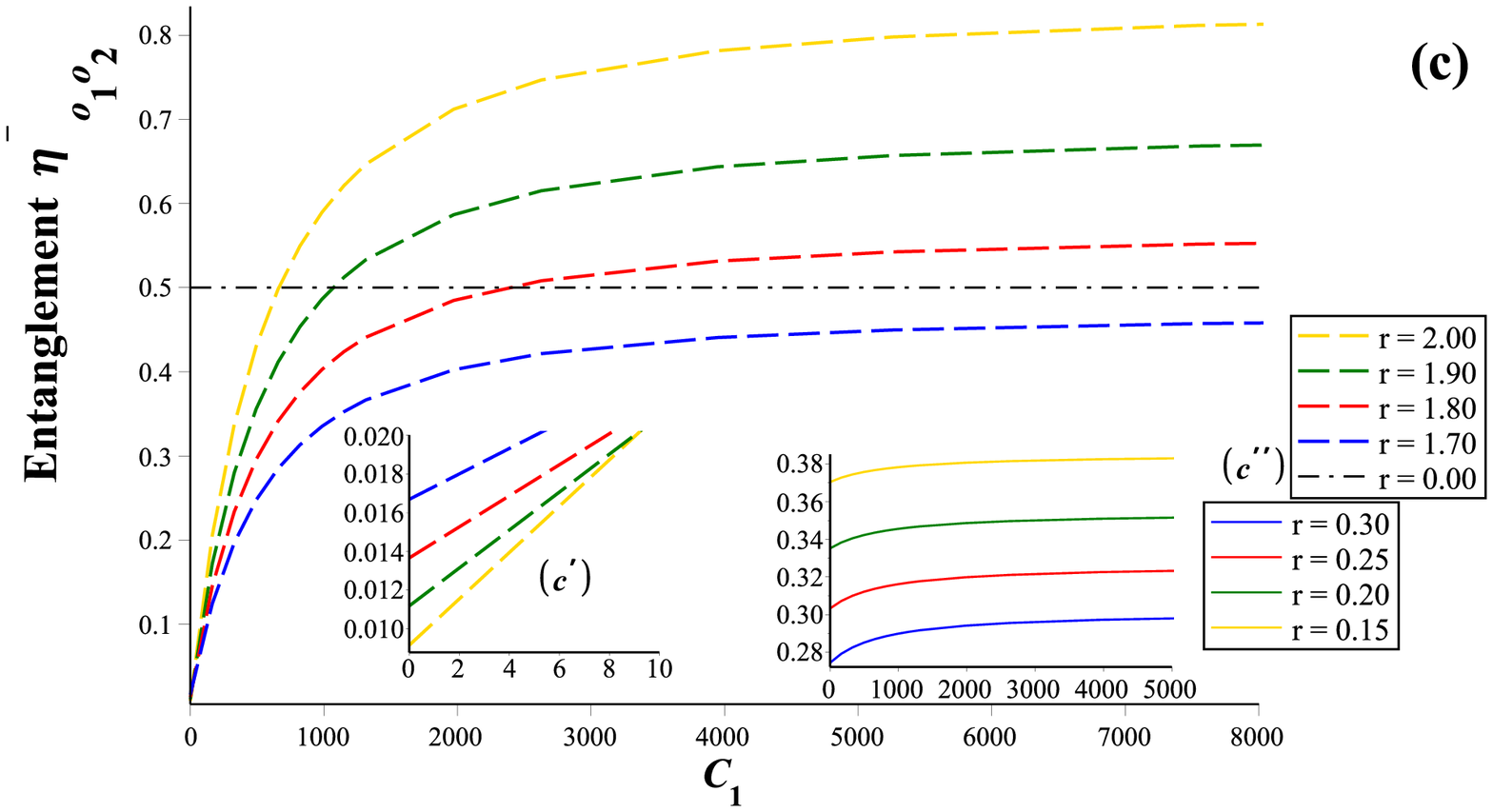}
\end{minipage}
} }
\caption{Plots of the three bipartite entanglements characterized by $%
\protect\eta _{j^{\prime }j^{\prime \prime }}^{-}$ ($j^{\prime }\neq
j^{\prime \prime }=o_{1},o_{2},m$) as functions of the dimensionless
optomechanical cooperativity $\mathcal{C}_{1}$ of the right cavity. $%
\mathcal{C}_{2}$ has been chosen to be $\mathcal{C}_{2}=2\mathcal{C}_{1}$
without loss of generality. Here we used $\protect\kappa _{1}=\protect\kappa %
_{2}=2\protect\pi \times 215\times 10^{3}~\mathrm{Hz}$, $\protect\gamma _{%
\protect\mu }=2\protect\pi .140~\mathrm{Hz}$ and $n_{\mathrm{th}}=10^{-2}$ ($%
T\simeq 9.\,\allowbreak 8~\protect\mu \mathrm{K}$). Different graphs
correspond to different values of the squeezing $r$. The panels (a), (b) and
(c) represent successively the entanglement between: the mechanical mode $m$
and the cavity mode $o_{1}$ ($\protect\eta _{mo_{1}}^{-}$), the mechanical
mode $m$ and the cavity mode $o_{2}$ ($\protect\eta _{mo_{2}}^{-}$) and
finally, the two optical cavities modes $o_{1}$ and $o_{2}$ ($\protect\eta %
_{o_{1}o_{2}}^{-}$). In the two insets, we are zooming into the region where 
$\mathcal{C}_{1}<50$. As seen in the inset ($c^{\prime \prime }$) of panel
(c), the purely optical subsystem remains always entangled for the same
squeezing values which have been used in the panels (a) and (b) regardless
the coupling regime. To switch from the entangled states to the others
separable in the purely optical subsystem, both broadband squeezed light and
strong coupling are required. For $r=0$ which corresponds to the horizontal
dotted-dashed line, the three subsystems are always separable whatever the
value of $\mathcal{C}_{1}$. }
\label{Fig.4}
\end{figure}
We further illustrate in Fig. \ref{Fig.4} the behavior of the three
functions $\eta _{_{mo_{1}}}^{-}$, $\eta _{_{mo_{2}}}^{-}$ and $\eta
_{_{o_{1}o_{2}}}^{-}$ with respect to the dimensionless optomechanical
cooperativity $\mathcal{C}_{1}$ ($\mathcal{C}_{2}$ $=2\mathcal{C}_{1}$) and
for various values of the squeezing parameter $r$. Here, we used $n_{\mathrm{%
th}}$ $=$ $10^{-2}$ (or equivalently $T\simeq 9.\,\allowbreak 8~\mu \mathrm{K%
}$) as value of the mean thermal photons number. The parameters $\gamma
_{\mu }$, $\kappa _{1}$ and $\kappa _{2}$ are the same as in Fig. \ref{Fig.3}%
. Focusing on the panels (a) and (b) in Fig. \ref{Fig.4}, it is clear that
in the case of weak coupling $\mathcal{C}_{1}\ll 1$ as well as in the strong
coupling $\mathcal{C}_{1}\gg 1$ (we note that, in the strong coupling limit, 
$\mathcal{C}_{1}$ can reach $10^{6}$ \cite{Clerk(4)}), the two
optomechanical hybrid subsystems are separable ($\eta _{_{mo_{1}}}^{-}>1/2$
and $\eta _{_{mo_{2}}}^{-}>1/2$). In contrast, for the same values of the
squeezing parameter $r$ which have been used in panels (a), (b) of Fig. \ref%
{Fig.4} (except $r=0$), the two optical modes $o_{1}$ and $o_{2}$ remain
always entangled regardless the coupling regime (see the inset ($c^{\prime
\prime }$) of panel (c) in Fig. \ref{Fig.4}). Whereas, to observe the
transition from the entangled states to the others separable in the purely
optical subsystem, both broadband squeezed light and strong coupling regime
are required. More interestingly, all the results obtained in Figs. \ref%
{Fig.3}, \ref{Fig.4} and \ref{Fig.5} show that, pumping the double-cavity
system by the squeezed light is a necessary condition to attained bipartite
and tripartite entanglement (for $r=0$, the three functions $\eta
_{_{mo_{1}}}^{-}$, $\eta _{_{mo_{2}}}^{-}$ and $\eta _{_{o_{1}o_{2}}}^{-}$
are always upper or equal $1/2$ regardless the others circumstance). This
can be interpreted as quantum correlations transfer from the two-mode
squeezed light to the three bipartite subsystems. According also to the
results obtained in Figs. \ref{Fig.3}, \ref{Fig.4} and \ref{Fig.5}, we have
shown that, in an experimentally accessible parameter regime, optomechanical
entanglement can be reached \textit{simultaneously} in the three bipartite
subsystems, confirming the existence of strong quantum correlations
distribution among the three modes. Finally, to close this section, we
emphasize that in various circumstances governed either by $T$, $r$ or $%
\mathcal{C}_{1}$, the largest and robust stationary bipartite entanglement
has been observed is the one between the two optical modes $o_{1}$ and $o_{2}
$ which are indirectly coupled.

\section{Quantum correlations beyond entanglement\label{sec4}}

\subsection{Gaussian quantum discord}

In this section, using the Gaussian quantum discord, we shall investigate
the non-classical correlations behavior in the three bipartite subsystems $%
m-o_{1}$, $m-o_{2}$ and $o_{1}-o_{2}$ far beyond entanglement. Recently, the
Gaussian quantum discord has been introduced to be more general than
entanglement as an indicator of non-classicality in (CV) systems. Indeed,
for same systems which exhibit non zero degree of mixture, the Gaussian
quantum discord can be non zero even at the separable state which is a
marker of the quantumness of correlations. Unlike entanglement, it has been
shown recently in a various publications that the Gaussian quantum discord
is more robust against dissipation and noise \cite{Olivares(1),Alonso} and
essentially, it does not undergo sudden death \cite%
{Mauro(1),IP,Olivares(1),Daoud}.

For a given bipartite Gaussian state, the analytical expression of the
Gaussian quantum discord is given by \cite{Adesso(2),Giorda} 
\begin{equation}
D^{j^{\prime }j^{\prime \prime }}=f\Big(\sqrt{\det B_{j^{\prime \prime }}}%
\Big)-f\Big(\nu _{+}^{j^{\prime }j^{\prime \prime }}\Big)-f\Big(\nu
_{-}^{j^{\prime }j^{\prime \prime }}\Big)+f\Big(\sqrt{\epsilon ^{j^{\prime
}j^{\prime \prime }}}\Big),\text{ for }j^{\prime }\neq j^{\prime \prime
}=o_{1},o_{2},m ,  \label{GQD}
\end{equation}
with $f(x)=\left( x+1/2\right) \ln \left( x+1/2\right) -\left( x-1/2\right)
\ln \left( x-1/2\right) $. Using the covariance matrix given by Eq. (\ref%
{Sub-CM}), we define the following five symplectic invariants \cite%
{Adesso(1)} $\alpha _{j^{\prime }}=\det B_{j^{\prime }},$ $\beta _{j^{\prime
\prime }}=\det B_{j^{\prime \prime }}$, $\theta _{j^{\prime }j^{\prime
\prime }}=\det C_{_{j^{\prime }j^{\prime \prime }}},$ $\lambda _{j^{\prime
}j^{\prime \prime }}=\det V_{_{j^{\prime }j^{\prime \prime }}}$ and $\tilde{%
\Delta}_{_{j^{\prime }j^{\prime \prime }}}=\alpha _{j^{\prime }}+\beta
_{j^{\prime \prime }}+2\theta _{j^{\prime }j^{\prime \prime }}$. The
nonpartially transposed symplectic eigenvalues $\nu _{+}^{j^{\prime
}j^{\prime \prime }}$ and $\nu _{-}^{j^{\prime }j^{\prime \prime }}$ which
are invariant under symplectic transformations are given by \cite%
{Vidal,Serafini}

\begin{equation}
\nu _{\pm }^{j^{\prime }j^{\prime \prime }}=\sqrt{\frac{\tilde{\Delta}%
_{_{j^{\prime }j^{\prime \prime }}}\pm \sqrt{\tilde{\Delta}_{_{j^{\prime
}j^{\prime \prime }}}^{2}-4\det V_{_{j^{\prime }j^{\prime \prime }}}}}{2}}.
\label{symplectic eigenvalues}
\end{equation}%
The explicit expression of the quantity $\epsilon ^{j^{\prime }j^{\prime
\prime }}$ which has been appeared in Eq. (\ref{GQD}) is defined by \cite%
{Adesso(2),Olivares(2)}.

\begin{equation}
\epsilon ^{j^{\prime }j^{\prime \prime }}=%
\begin{cases}
\frac{2(\theta _{j^{\prime }j^{\prime \prime }})^{2}+(1/4-\beta _{j^{\prime
\prime }})(\alpha _{j^{\prime }}-4\lambda _{j^{\prime }j^{\prime \prime
}})+2|\theta _{j^{\prime }j^{\prime \prime }}|\sqrt{(\theta _{j^{\prime
}j^{\prime \prime }})^{2}+(1/4-\beta _{j^{\prime \prime }})(\alpha
_{j^{\prime }}-4\lambda _{j^{\prime }j^{\prime \prime }})}}{4(1/4-\beta
_{j^{\prime \prime }})^{2}}\text{ \ \ \ if \ \ }d^{j^{\prime }j^{\prime
\prime }}\leq 0 \\ 
\frac{\alpha _{j^{\prime }}\beta _{j^{\prime \prime }}-(\theta _{j^{\prime
}j^{\prime \prime }})^{2}+\lambda _{j^{\prime }j^{\prime \prime }}-\sqrt{%
(\theta _{j^{\prime }j^{\prime \prime }})^{4}+(\lambda _{j^{\prime
}j^{\prime \prime }}-\alpha _{j^{\prime }}\beta _{j^{\prime \prime
}})^{2}-2(\theta _{j^{\prime }j^{\prime \prime }})^{2}(\alpha _{j^{\prime
}}\beta _{j^{\prime \prime }}+\lambda _{j^{\prime }j^{\prime \prime }})}}{%
2\beta _{j^{\prime \prime }}}\text{ \ \ if \ }d^{j^{\prime }j^{\prime \prime
}}>0%
\end{cases}%
,  \label{Epsilon}
\end{equation}%
where the discriminant $d^{j^{\prime }j^{\prime \prime }}$ is given by 
\begin{equation}
d^{j^{\prime }j^{\prime \prime }}=\Big(\lambda _{j^{\prime }j^{\prime \prime
}}-\alpha _{j^{\prime }}\beta _{j^{\prime \prime }}\Big)^{2}-\Big(1/4+\beta
_{j^{\prime \prime }}\Big)\Big(\theta _{j^{\prime }j^{\prime \prime }}\Big)%
^{2}\Big(\alpha _{j^{\prime }}+4\lambda _{j^{\prime }j^{\prime \prime }}\Big)%
.  \label{g_discriminant}
\end{equation}%
In Eq. (\ref{GQD}), the term $f\Big(\sqrt{\det B_{j^{\prime \prime }}}\Big)$
is the Von-Neumann entropy of the reduced state of the $j^{\prime \prime }$%
-mode. Whereas, the quantity $f\Big(\nu _{+}^{j^{\prime }j^{\prime \prime }}%
\Big)+f\Big(\nu _{-}^{j^{\prime }j^{\prime \prime }}\Big)$ is the entropy of
the bipartite subsystem formed by the two modes $j^{\prime}$ and $j^{\prime
\prime}$. On the other hand, $f\Big(\sqrt{\epsilon ^{j^{\prime}j^{\prime
\prime }}}\Big)$ represents the entropy of the $j^{\prime }$-mode after
performing a Gaussian measurement on the $j^{\prime \prime }$-mode, where
the measurement is chosen to minimize this quantity. In what follows, we
shall denote by $D^{mo_{1}}$ (respectively. $D^{mo_{2}}$) the Gaussian
quantum discord between the mechanical mode $m$ and the optical mode $o_{1}$
(respectively. $o_{2}$). Similarly, $D^{o_{1}o_{2}}$ stands for the Gaussian
quantum discord between the two optical modes $o_{1}$ and $o_{2}$.

\subsection{ Non-classical correlations far beyond entanglement}

\textbf{\textit{4.2.1~~Gaussian quantum discord versus the thermal effect}} 
\newline

\begin{figure}[tbh]
\textrm{\textrm{\centering
\begin{minipage}[htb]{2in}
\centering
\includegraphics[width=2in]{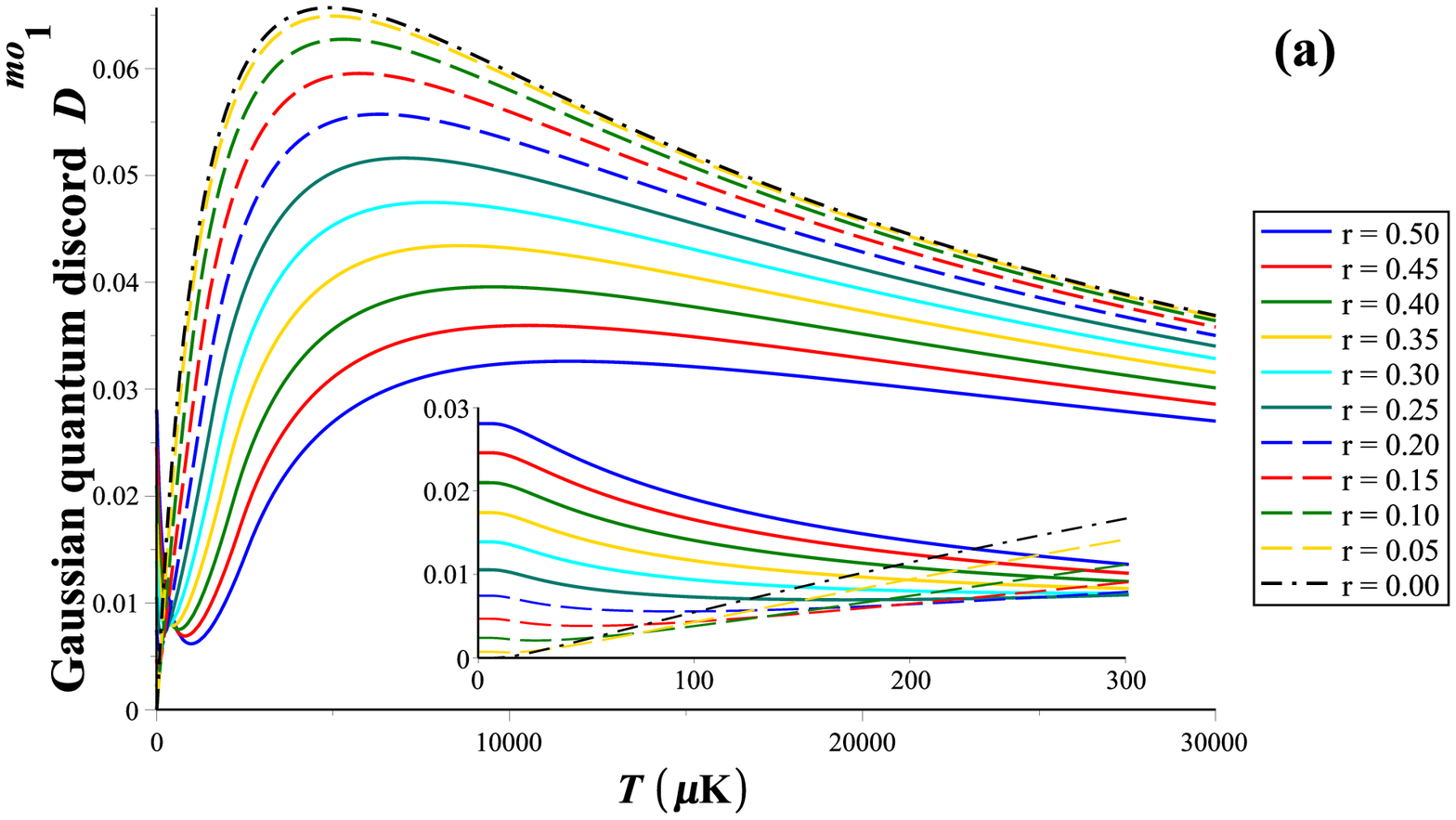}
\end{minipage}
\begin{minipage}[htb]{2in}
\centering
 \includegraphics[width=2in]{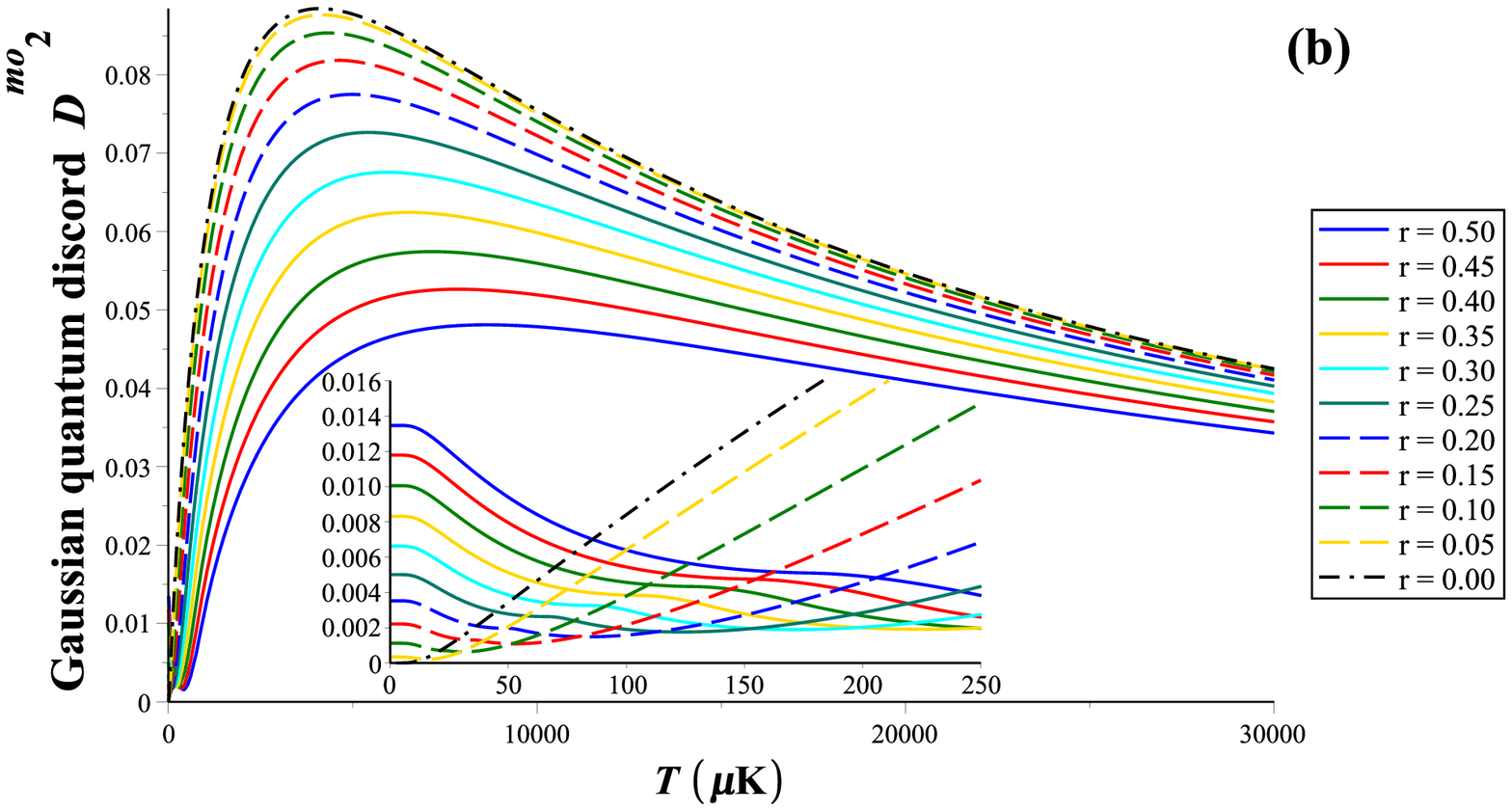}
\end{minipage}
\begin{minipage}[htb]{2in}
\centering
 \includegraphics[width=2in]{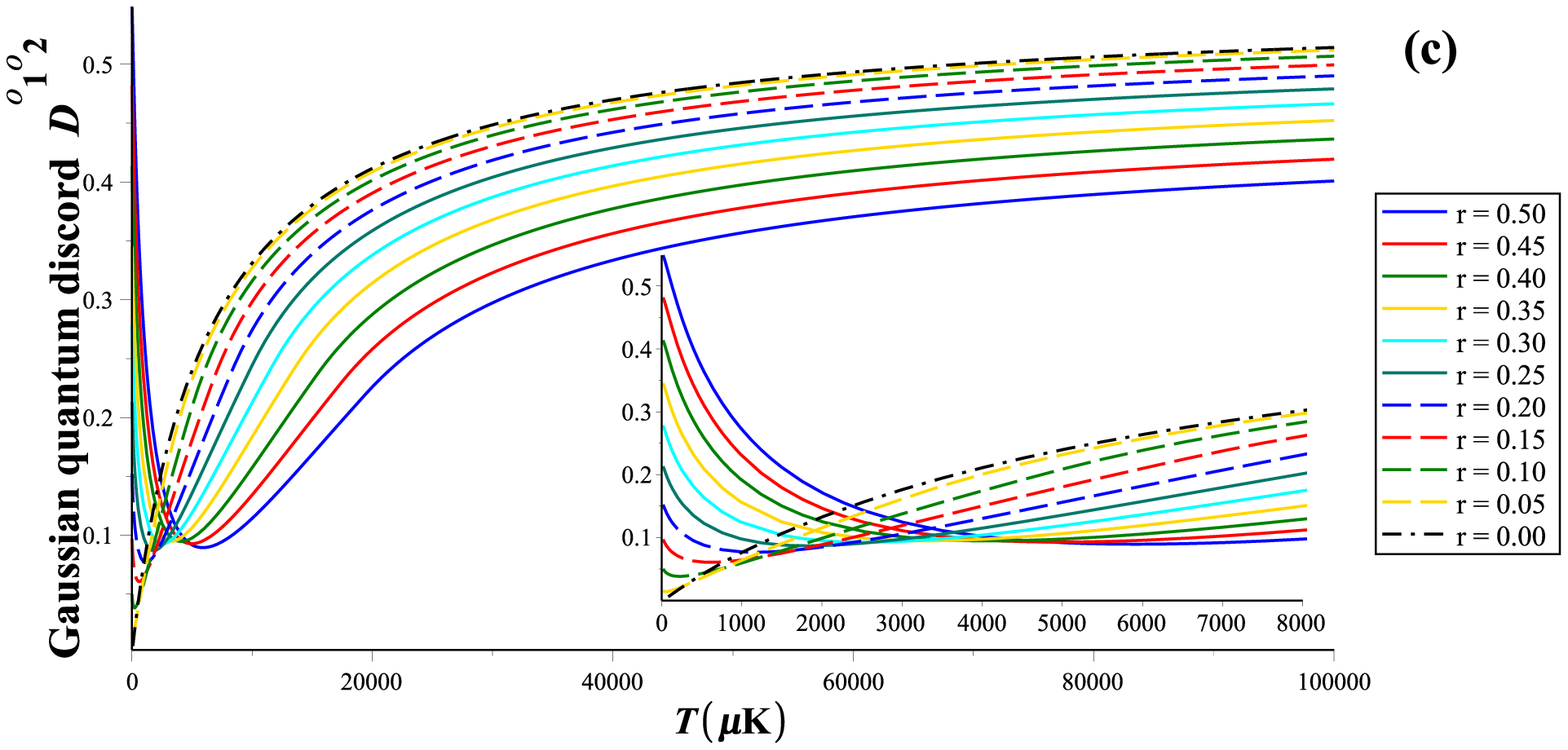}
\end{minipage}
} }
\caption{ Plots of the three Gaussian quantum discords $D^{mo_{1}}$ (panel
(a)), $D^{mo_{2}}$ (panel (b)) and $D^{o_{1}o_{2}}$ (panel (c)) versus the
thermal bath temperature $T$ of the movable mirror and for various values of
the squeezing parameter $r$. The parameters are the same as in Fig. \protect
\ref{Fig.2}. The nonzero asymptotic values of $D^{mo_{1}}$, $D^{mo_{2}}$ and 
$D^{o_{1}o_{2}}$ which corresponding to the case where the whole system is
fully separable (see Fig. \protect\ref{Fig.2} where $\protect\eta %
_{mo_{1}}^{-}>1/2$, $\protect\eta _{mo_{2}}^{-}>1/2$ and $\protect\eta %
_{o_{1}o_{2}}^{-}>1/2$ for a given squeezing $r$), confirm the existence of
the simultaneous quantumness of correlations in the three subsystems.
Non-classicality of the two hybrid subsystems (respectively. the purely
optical subsystem) can be observed for temperatures more than $T=0.03K$
(respectively. $T=0.1K$). Interestingly enough, the simultaneous reduction
of $D^{mo_{1}}$ and $D^{mo_{2}}$ from $T\approx5m\mathrm{K}$ is clearly
accompanied by the enhancement of $D^{o_{1}o_{2}}$. These opposites
behaviors can be explained by the purely quantum correlations transfer from
the two hybrid subsystems to the homogeneous optical subsystem by means of
the movable mirror.}
\label{Fig.5}
\end{figure}
The robustness of the different pairwise Gaussian quantum discord on the
mechanical thermal bath temperature $T$ and for various values of the
squeezing parameters $r$ is illustrated in Fig. \ref{Fig.5}. In order to
compare the behavior of the entanglement (Fig. \ref{Fig.2}) and the Gaussian
quantum discord (Fig. \ref{Fig.5}) under the thermal effects, we have fixed
the parameters as the same as in (Fig. \ref{Fig.2}). The panels (a), (b) and
(c) show that the three Gaussian quantum discords $D^{mo_{1}}$, $D^{mo_{2}}$
and $D^{o_{1}o_{2}}$ remain non zero whatever the value of $r$ as well as
for high values of the temperature $T$. Next, comparing the results which
illustrated in Figs. \ref{Fig.2} and \ref{Fig.5}, it is clear that unlike
the three bipartite entanglements $m-o_{1}$, $m-o_{2}$ and $o_{1}-o_{2}$,
the three Gaussian quantum discords $D^{mo_{1}}$, $D^{mo_{2}}$ and $%
D^{o_{1}o_{2}}$ do not undergo the sudden death and remain non zero even for
large values of the temperatures $T$ as expected. This reflects the robust
character of the Gaussian quantum discord against strong thermal noise. More
important, non-classicality of the two hybrid subsystems is significantly
nonzero and persists for temperatures up to $0.03 $\textrm{K}. Whereas, the
quantumness of correlations can be detected more than $T$ $=0.1$\textrm{K}
in the purely optical subsystem, which is almost of the same order of
magnitude that has been observed in Ref \cite{Mauro(1)} for a purely
mechanical system. Interestingly enough, Fig. \ref{Fig.5} shows that, the
Gaussian quantum discords $D^{mo_{1}}$ and $D^{mo_{2}}$ start to decay
asymptotically with $T$ from $T\approx5m\mathrm{K} $ (see panels (a) and (b)
in Fig. \ref{Fig.5}). On the other hand, the amount of the Gaussian quantum
discord $D^{o_{1}o_{2}}$ monotonically increases with $T $ also from $%
T\approx5m\mathrm{K}$ (see panel (c) in Fig. \ref{Fig.5}), which is a
surprising quantum correlations behavior against thermal effects. This
result can be explained quantitatively by the purely quantum correlations
transfer from the two hybrid subsystems to the homogeneous optical by
mediation of the movable mirror shared between the two cavities. \newline

\textbf{\textit{4.2.2~~Gaussian quantum discord versus the squeezing effect}}
\newline

\begin{figure}[tbh]
\textrm{\textrm{\centering
\begin{minipage}[htb]{2in}
\centering
\includegraphics[width=2in]{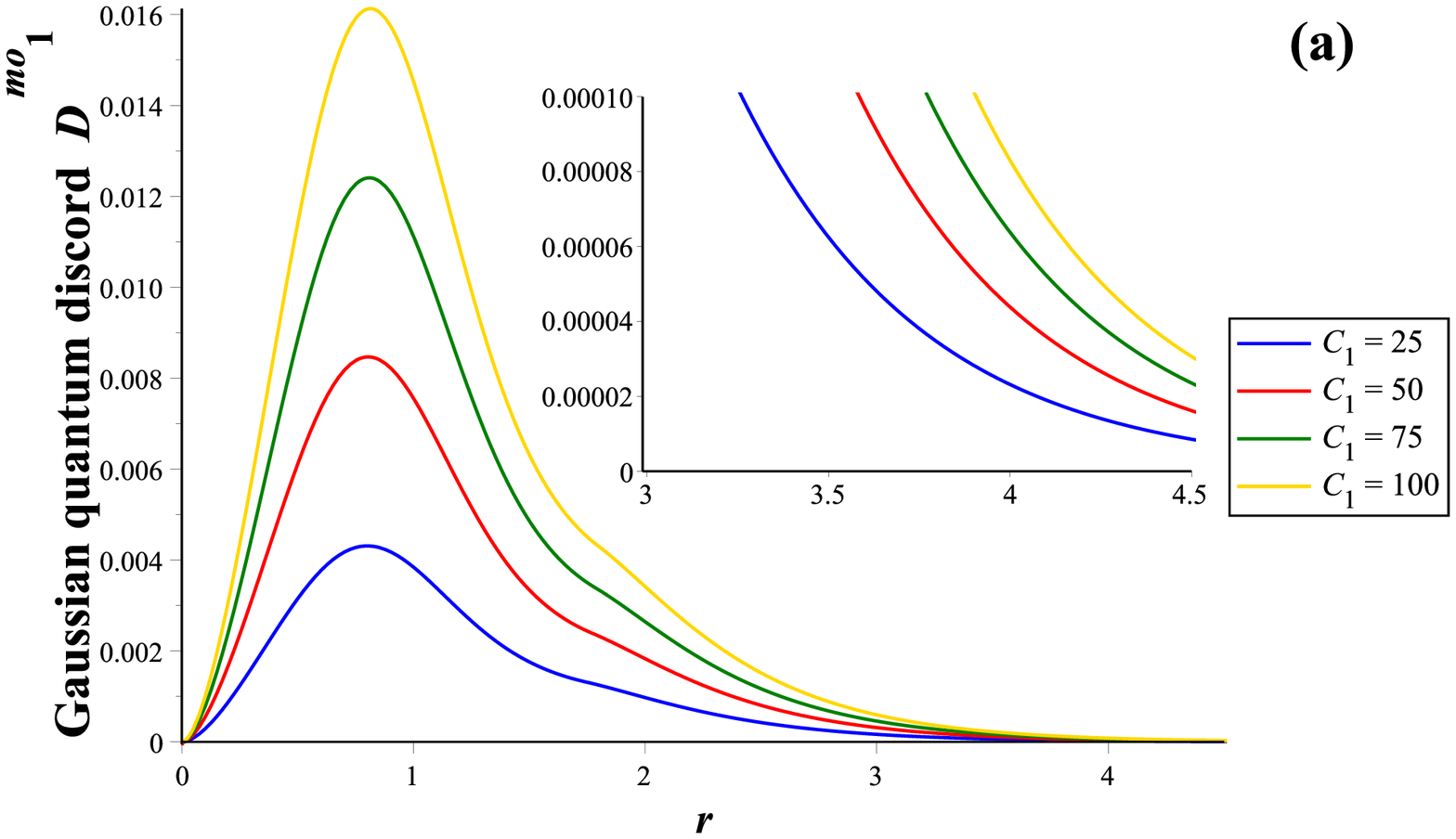}
\end{minipage}
\begin{minipage}[htb]{2in}
\centering
 \includegraphics[width=2in]{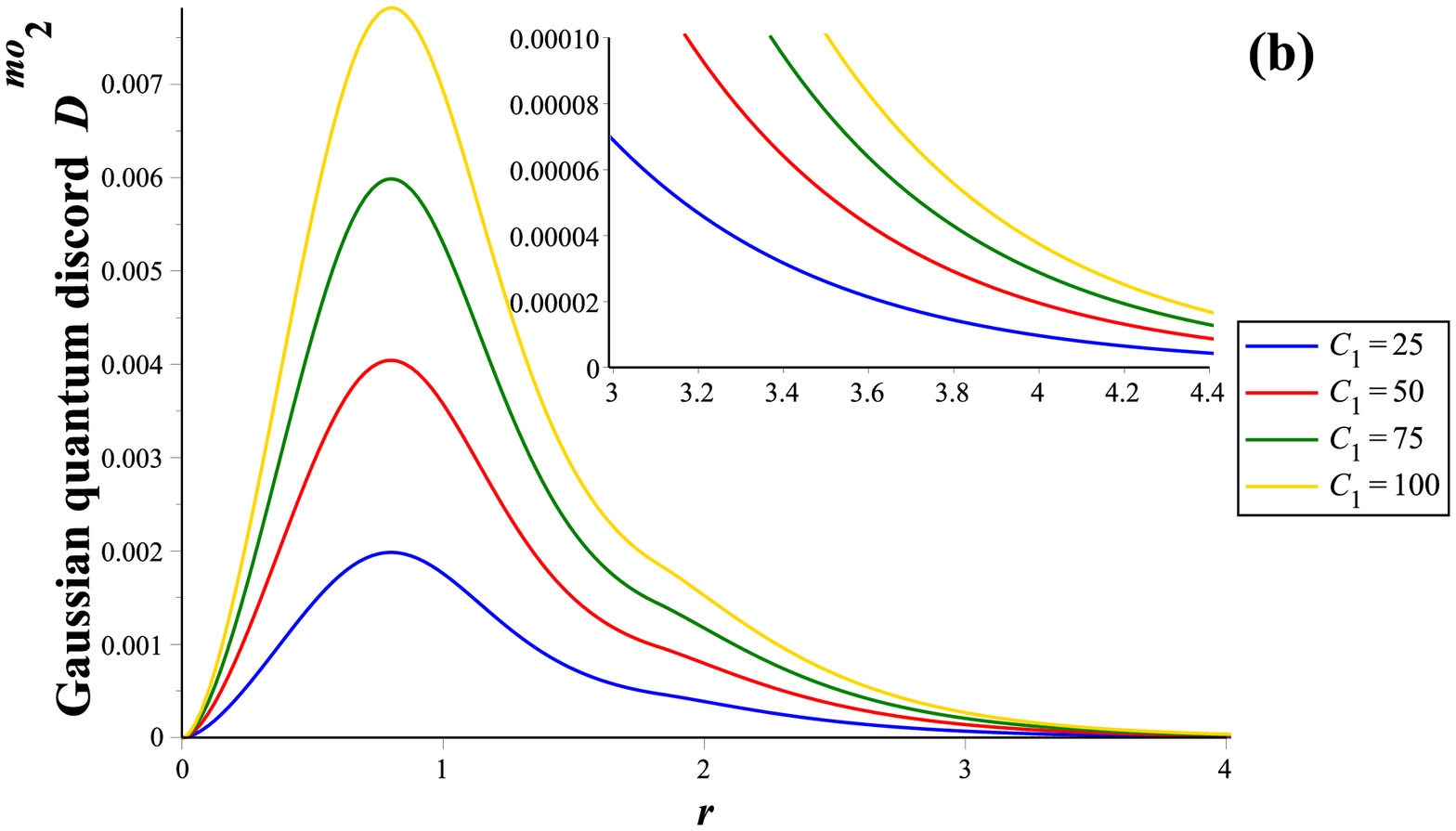}
\end{minipage}
\begin{minipage}[htb]{2in}
\centering
 \includegraphics[width=2in]{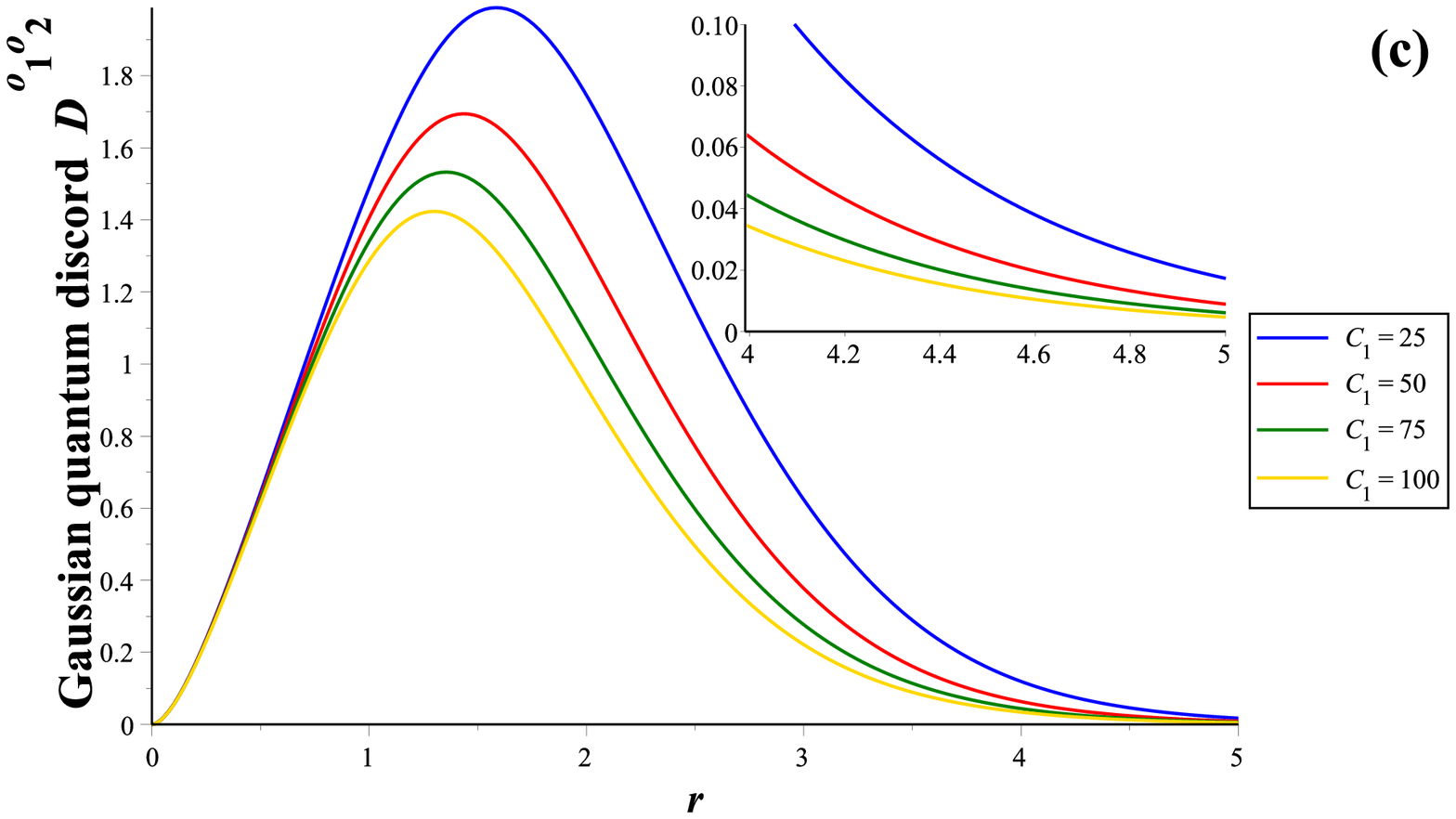}
\end{minipage}
} }
\caption{ The three Gaussian quantum discords $D^{mo_{1}}$ (panel (a)), $%
D^{mo_{2}}$ (panel (b)) and $D^{o_{1}o_{2}}$ (panel (c)) against the
squeezing parameter $r$ and for different values of the optomechanical
cooperativity $\mathcal{C}_{1}$ ($\mathcal{C}_{2}=2\mathcal{C}_{1}$). The
parameters are the same as in Fig. \protect\ref{Fig.3}. The three functions $%
D^{mo_{1}}$, $D^{mo_{2}}$ and $D^{o_{1}o_{2}}$ exhibit a resonance like
behavior with respect to the squeezing $r$. Comparing the panels (a), (b)
and (c) in this figure with their corresponding in Fig. \protect\ref{Fig.3},
we can see clearly that when the three subsystems are completely separable ($%
\protect\eta _{mo_{1}}^{-}>1/2$, $\protect\eta _{mo_{2}}^{-}>1/2$ and $%
\protect\eta _{o_{1}o_{2}}^{-}>1/2$ for a given value of $\mathcal{C}_{1}$),
non-classical correlations can be observed simultaneously in the three
subsystems ($D^{mo_{1}}>0$, $D^{mo_{2}}>0$ and $D^{o_{1}o_{2}}>0$) even for
squeezing values more than $r=4.5$ (see the insets in Fig. \protect\ref%
{Fig.6}).}
\label{Fig.6}
\end{figure}
The influence of the squeezing on the three Gaussian quantum discords $%
D^{mo_{1}}$, $D^{mo_{2}}$ and $D^{o_{1}o_{2}}$ and for different amounts of
the dimensionless optomechanical cooperativity $\mathcal{C}_{1}$ is
illustrated in Fig. \ref{Fig.6}. The plots are done by fixing the parameters
similarly to ones used to obtain the results reported in Fig. \ref{Fig.3}.
As seen in Fig. \ref{Fig.6}, the three Gaussian quantum discords $D^{mo_{1}}$%
, $D^{mo_{2}}$ and $D^{o_{1}o_{2}}$ exhibit a resonance like behavior with
respect to the squeezing parameter $r$ for a given value of $\mathcal{C}_{1}$%
. This interesting result means that by controlling the level of the
squeezing, one can reach the situation where the two considered modes are
maximally discordant. Furthermore, we remark also that when the whole system
is straightforwardly separable (see for example the case in Fig. \ref{Fig.3}
which corresponds to $\mathcal{C}_{1}=25$ and $r>3.5$), the three Gaussian
quantum discords $D^{mo_{1}}$, $D^{mo_{2}}$ and $D^{o_{1}o_{2}}$ are
asymptotically non-zero. This reflects the simultaneous existence of the
quantumness of correlations in the three optomechanical subsystems even for
squeezing values upper to $r=4.5$ (see the three insets in Fig. \ref{Fig.6}).%
\newline

\textbf{\textit{4.2.3~~Gaussian quantum discord versus the optomechanical
coupling}} \newline

\begin{figure}[tbh]
\textrm{\textrm{\centering
\begin{minipage}[htb]{2in}
\centering
\includegraphics[width=2in]{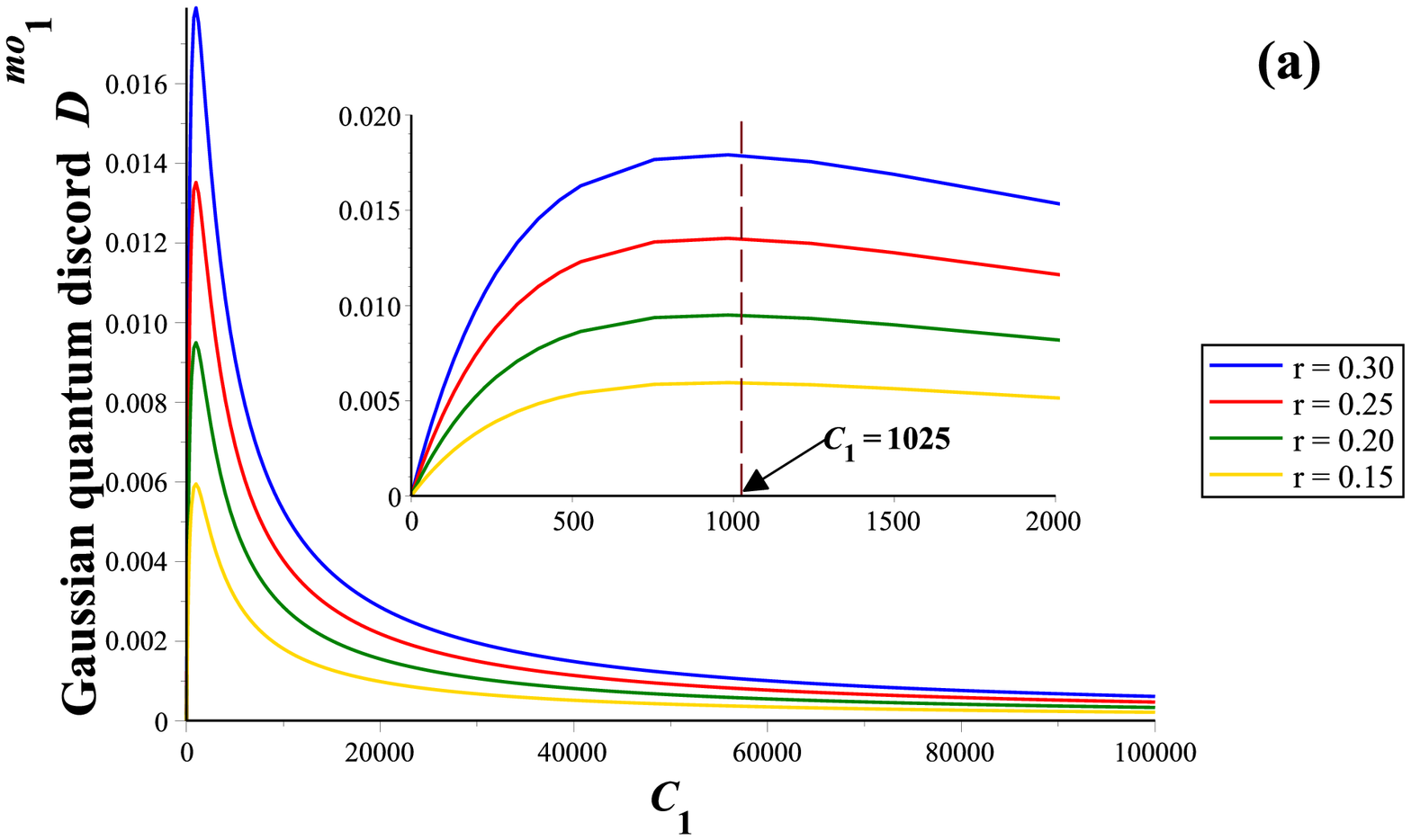}
\end{minipage}
\begin{minipage}[htb]{2in}
\centering
 \includegraphics[width=2in]{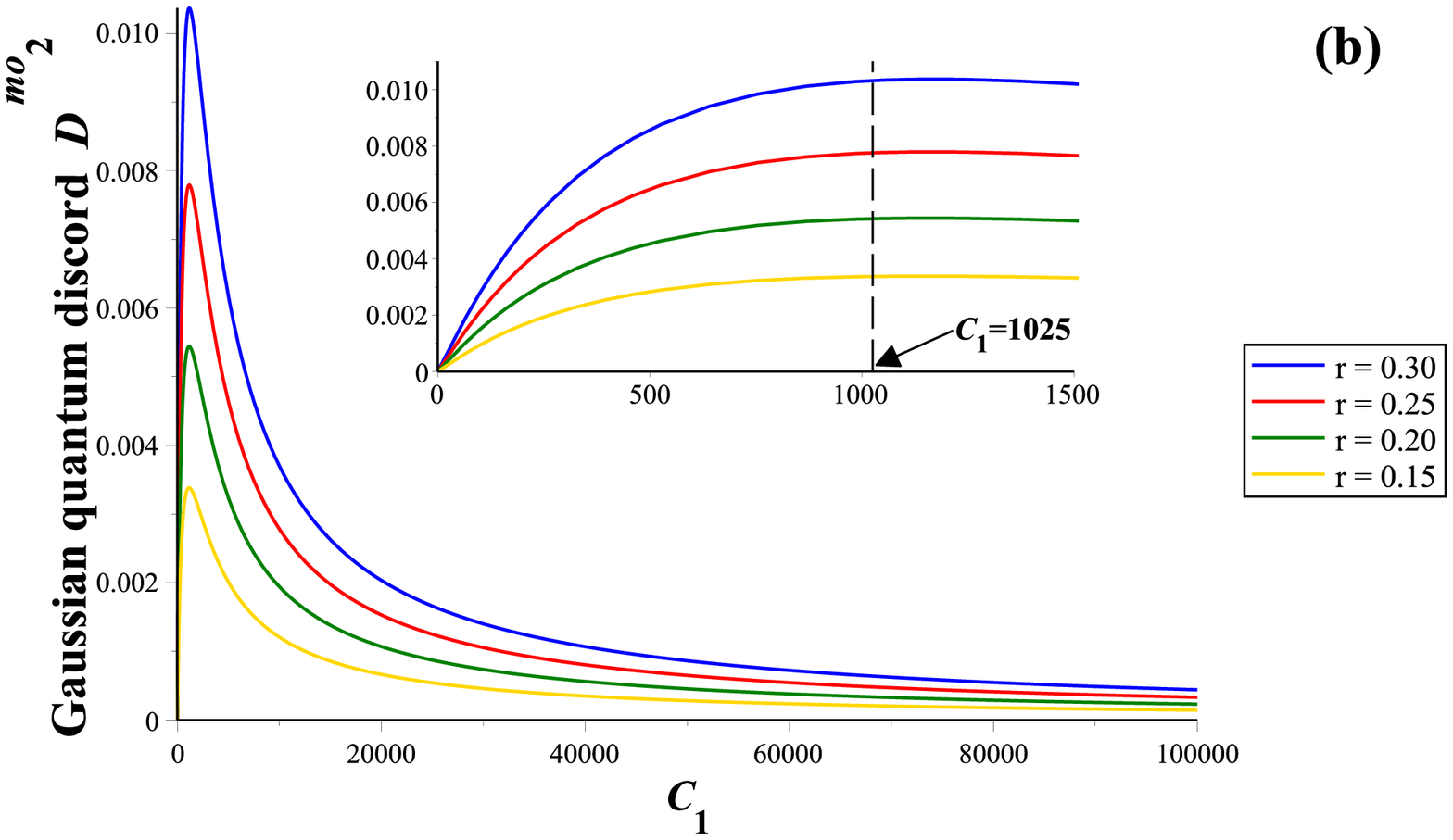}
\end{minipage}
\begin{minipage}[htb]{2in}
\centering
 \includegraphics[width=2in]{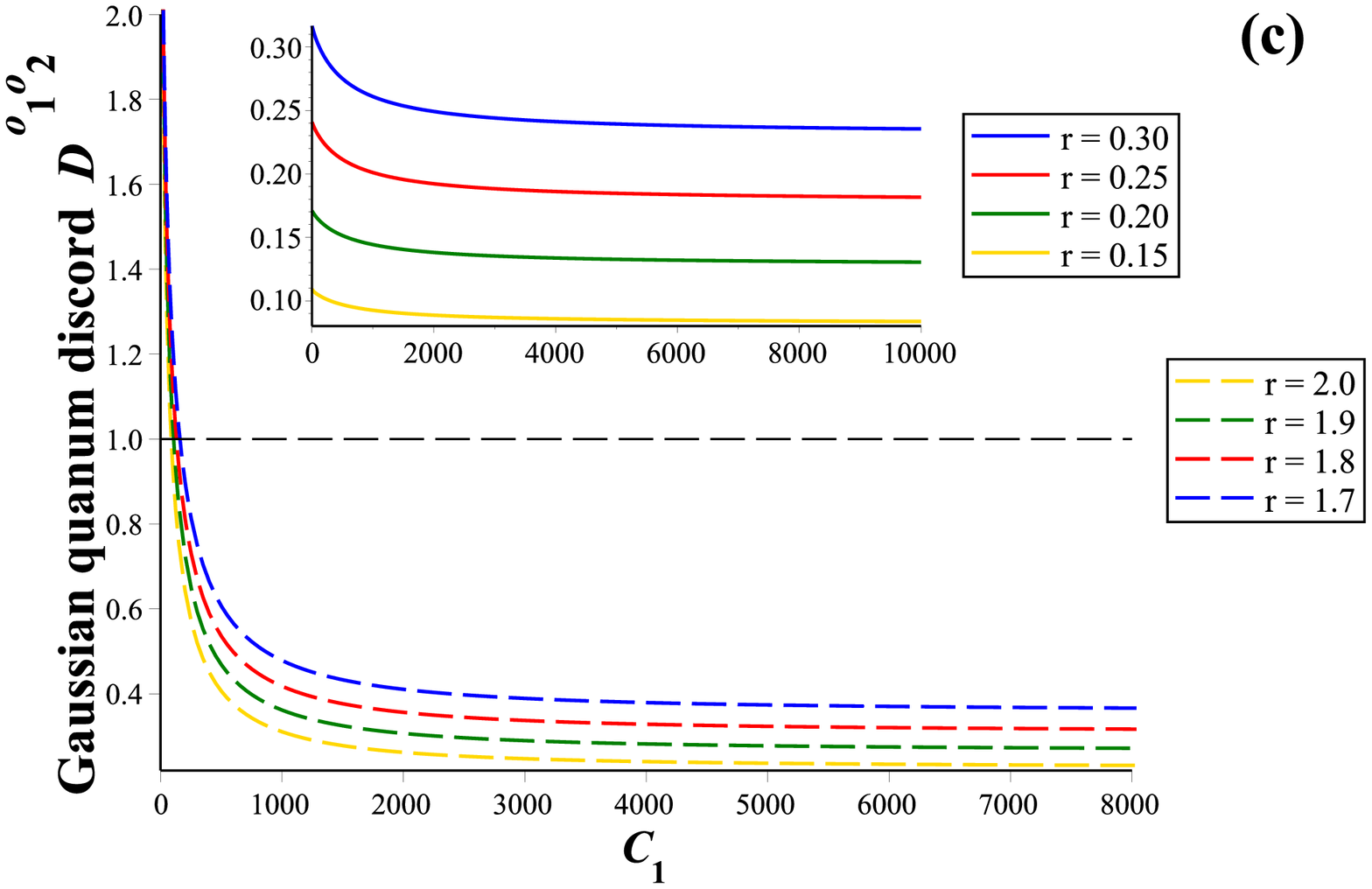}
\end{minipage}
} }
\caption{ The three Gaussian quantum discords $D^{mo_{1}}$ (panel (a)), $%
D^{mo_{2}}$ (panel (b)) and $D^{o_{1}o_{2}}$ (panel (c)) against the
dimensionless optomechanical cooperativity $\mathcal{C}_{1}$ ($\mathcal{C}%
_{2}=2\mathcal{C}_{1}$). Different values of the squeezing parameter $r$
have been used. The parameters are the same as in Fig. \protect\ref{Fig.4}.
For a given squeeze parameter $r$, the situation where $\protect\eta %
_{j^{\prime }j^{\prime \prime }}^{-}>1/2$ (see Fig. \protect\ref{Fig.4}) and 
$D^{j^{\prime }j^{\prime \prime }}>0$ (with $j^{\prime }\neq j^{\prime
\prime }=o_{1},o_{2},m$), witnesses the existence of the quantumness of
correlations in the state of the two considered modes $j^{\prime }$ and $%
j^{\prime \prime }$.}
\label{Fig.7}
\end{figure}

Finally, we study the dependence of the three Gaussian quantum discords $%
D^{mo_{1}}$, $D^{mo_{2}}$ and $D^{o_{1}o_{2}}$ with the dimensionless
optomechanical cooperativity $\mathcal{C}_{1}$. The resulting behavior is
presented in the panels (a), (b) and (c) of Fig. \ref{Fig.7} for various
amounts of the squeezing parameters $r$. 
As illustrated in Fig. \ref{Fig.7}, for a given squeeze parameter $r$, the
three bipartite Gaussian quantum discords ${D}^{mo_{1}}$, ${D}^{mo_{2}}$ and 
${D}^{o_{1}o_{2}}$ are always non-zero (except $\mathcal{C}_{1}=0$ in panels
(a) and (b)) regardless the coupling regime. It is also remarkable that in
the two hybrid subsystems, the corresponding Gaussian quantum discord ${D}%
^{mo_{1}}$ and ${D}^{mo_{2}}$ are always maximal around $\mathcal{C}%
_{1}\approx 1025$ whatever the squeezing parameter $r$ (see the insets of
panels (a) and (b) in Fig. \ref{Fig.7}). It follows that one can reach the
maximum amount of the Gaussian quantum discord in the two hybrid subsystems
by a judicious tuning of the physical parameters determining the
optomechanical cooperativity (see Eq. (\ref{OCooperativity})). Finally,
comparing the behavior of the $j^{\prime }-j^{\prime \prime }$ entanglement
quantified by $\eta _{j^{\prime }j^{\prime \prime }}^{-}$ and the
corresponding Gaussian quantum discord $D^{j^{\prime }j^{\prime \prime }}$
(with $j^{\prime }\neq j^{\prime \prime }=o_{1},o_{2},m$) under the effect
of either $T$, $r$ or $\mathcal{C}_{1}$, we remark that: \textit{(i)} for
regions where the Gaussian quantum discord $D^{j^{\prime }j^{\prime \prime
}} $ is greater than $1$, the two modes $j^{\prime }$ and $j^{\prime \prime
} $ are entangled ($\eta _{j^{\prime }j^{\prime \prime }}^{-}<1/2$) and 
\textit{(ii)} the Gaussian quantum discord can be less than $1$ for both
entangled and separable states of the two considered modes $j^{\prime }$ and 
$j^{\prime \prime }$, which is in agreement with the proprieties of the
Gaussian quantum discord \cite{Adesso(2),Giorda}.

\section{Concluding Remarks}

To summarize, both entanglement and Gaussian quantum discord have been
studied in a tripartite optomechanical setup (fed by squeezed light)
comprising two optical cavities modes and a single mechanical mode. Using a
linearized fluctuations analysis under the Markovian process, the $6\times 6$
covariance matrix encoding the essential of the correlations between the
different modes, is derived in the resolved-sideband regime. For
entanglement characterization, we used the Simon criterion. In order to
capture the quantumness of correlations, the Gaussian quantum discord is
evaluated employing the method reported in \cite%
{Adesso(2),Giorda,Olivares(2)}. The entanglement and the Gaussian quantum
discord of three different bipartite subsystems have been evaluated as
functions of the thermal bath temperature $T$, the squeezing parameter $r$
and also the optomechanical coupling $\mathcal{C}_{1}$. For an
experimentally accessible parameter regime, we showed that it is possible to
transfer the quantum correlations from the two-mode squeezed light to the
system, creating simultaneous three bipartite optomechanical entanglements.
It was further seen that under different circumstance, the purely optical
subsystem exhibits more intricacy in comparison with the two hybrid
subsystems. As expected, the three bipartite Gaussian quantum discords are
shown more resilient against the destructive effects. More important, they
are always nonzero even for extremal limiting situations as higher
temperature $T$, large value of the squeezing $r$ and also for both strong
coupling ($\mathcal{C}_{1}\gg1$) and weak coupling ($\mathcal{C}_{1}\ll1$).
In general, under various conditions, the quantumness of correlations has
been detected simultaneously in the three bipartite subsystems over a very
wide range of the parameters characterizing the environment and the system.
Finally, one should recognize that our study in this two-mode optomechanical
system is limited to pairwise quantum correlations. This is essentially due
to the fact the characterization of entanglement in multipartite quantum
systems remains a complex and open issue. However, we think that the
tripartite entanglement classification given in Ref \cite{Giedke(2)}, which
provides a necessary and sufficient criterion based on the nonpositive
partial transposition \cite{RFWerner}, can be used as an alternative way to
deal with entanglement in the case of tripartite continuous variables
Gaussian states. Also, it will be important to consider the balance of
pairwise quantum correlations in the tripartite optomechanical system
considered here in the spirit of the ideas developed recently in \cite%
{M.G.Paris(1)}. We hope to report on these issues in a forthcoming works. 
\newline

\end{document}